\documentclass[%
 reprint,
showpacs,
preprintnumbers,
 amsmath,
 amssymb,
 aps,
 prd,
]{revtex4-2}

\usepackage{graphicx}
\usepackage{dcolumn}
\usepackage{bm}

\begin{document}

\title{Scattering of fermionic isodoublets on the sine-Gordon kink}

\author{A.~Yu.~Loginov}

\email{a.yu.loginov@tusur.ru}

\affiliation{Tomsk State University of Control Systems and Radioelectronics, 634050 Tomsk, Russia}


\date{\today}

\begin{abstract}
The scattering of  Dirac  fermions  on  the  sine-Gordon  kink  is studied both
analytically and numerically.
To achieve invariance  with  respect  to  a  discrete symmetry, the sine-Gordon
model is treated as a  nonlinear  $\sigma$-model  with  a circular target space
that interacts with fermionic isodublets through the Yukawa interaction.
It is shown that the diagonal  and  antidiagonal  parts  of  the fermionic wave
function interact independently with the external field of the sine-Gordon kink.
The wave functions of the fermionic scattering states are expressed in terms of
the Heun functions.
General  expressions  for  the  transmission  and  reflection  coefficients are
derived, and their dependences  on  the  fermion  momentum and mass are studied
numerically.
The existence condition  is  found  for  two  fermionic  zero  modes, and their
analytical expressions are obtained.
It is shown that the zero modes do not lead to fragmentation of  the  fermionic
charge, but can lead to polarization of the fermionic vacuum.
The scattering of the diagonal and antidiagonal fermionic states is found to be
significantly different; this difference is shown to be due  to  the  different
dependences of the energy levels of these bound states on the fermion mass, and
is  in accordance with Levinson’s theorem.
\end{abstract}

\maketitle

\section{\label{sec:I} Introduction}

Many field theory models  with  spontaneously  broken  symmetry  possess stable
localized solutions known as topological solitons \cite{Manton}.
One example of a topological soliton   is   a  kink \cite{Vachaspati}, which is
a one-dimensional static solution of a $(1+1)$-dimensional field model.
The best known are the kinks of the $\phi^{4}$ \cite{dhr_74, polyakov_jetpl_74,
 goldstone_jackiw_prd_75}  and  the  sine-Gordon  models  \cite{skyrme_prsl_61,
  skyrme_np_62}.
The sine-Gordon model has a number of remarkable  properties; in particular, it
possesses an infinite number of conserved quantities  at both the classical and
quantum levels.
Furthermore, all time-dependent solutions  of  the sine-Gordon model are known,
and can be written down explicitly in closed form.
The reason for this is that the sine-Gordon  model  in $(1+1)$ dimensions is an
integrable system.
Another special  property  of  the  sine-Gordon model is its equivalence to the
massive Thirring model, and the quantum soliton of the sine-Gordon model can be
identified with the fermion of the massive Thirring model \cite{colemal_prd_75}.

The sine-Gordon model and its modifications  are  used  in  the study of a wide
range  of  phenomena, including  QCD  \cite{blas_jhep01_2007, blas_jhep03_2007,
 nitta_prd_2013}, condensed  matter  physics  \cite{Barone},  and  solid  state
physics in relation to  the Josephson junctions and their associated  magnetism
and topological excitations \cite{ddkp_prl_85, gwtu_prb_1998, njnsvm_jpcm_1998,
 nvmjns_jpcm_1998}.
Models of the  sine-Gordon  type  are  also  used to describe  vortex  dynamics
in superconducting systems \cite{bcg_prl_2007} and to investigate the Josephson
current in some systems \cite{css_prl_2002}.

The interaction  between  fermions  and  the  background  fields of topological
solitons  leads  to  a  number  of   interesting   physical  effects,  such  as
fragmentation of the fermionic number and polarization of  the fermionic vacuum
\cite{jackiw_rebbi_prd_76, goldstone_wilczek_prl_81,  mackenzie_wilczek_prd_84,
 niemi_semenoff_86},  superconductivity   \cite{witten_npb_85},  and   monopole
catalysis   of   the   proton   decay   \cite{rubakov_jetpl_81, rubakov_npb_82,
 callan_prd_25_1982, callan_prd_26_1982}.
In the case of a  kink, fragmentation of the fermionic number  and polarization
of the fermionic vacuum are possible \cite{jackiw_rebbi_prd_76}.
In addition, the kink's  field  distorts  the  energy  levels  of the fermionic
vacuum; bound states  can  arise  and continuum states can change compared to a
free fermion.
These lead to a change in the  zero-point  fermion  energy, and consequently to
the Casimir effect,  in  the  presence  of  the  kink \cite{gousheh_EPJC_2014}.

The fermion-kink   interactions,   kink-antikink   configurations,  and  domain
walls   were   also  considered   in   Refs.~\cite{voloshin,  ayala,  funakubo,
farrar1,farrar2,stojkovic_prd_00,campanelli1,campanelli2,chu_vachaspati_prd_08,
 brih_prd_2008, loginov_prd_2017, perapechka_prd_2020}.
The main purpose of the present  work  is  to  study the  scattering  and bound
states of fermions in the external field of the sine-Gordon kink.
A characteristic property  of  the  sine-Gordon  model  is  its invariance with
respect to discrete equidistant shifts of a scalar field.
The usual Yukawa interaction of a fermion  with  the  scalar  field  will break
this discrete $\mathbb{Z}$-symmetry  of  the sine-Gordon model, and in order to
avoid this,  the  sine-Gordon  model  is  treated as a nonlinear $\sigma$-model
whose target space is a circle, and  the  scalar field is treated as an angular
variable.
Under  these  conditions,  the   Yukawa  interaction  of  the  $\sigma$-model's
two-component nonlinear scalar field  with  a  fermionic  isodoublet  will  not
break  the  discrete $\mathbb{Z}$-symmetry.

This  paper  is  structured as follows.
In Sec.~\ref{sec:II}, we describe  briefly  the  Lagrangian,  symmetries, field
equations, and  kink  solution  of the sine-Gordon model.
In  Sec.~\ref{sec:III},   we   consider  fermion-kink  scattering,  and  derive
analytical  expressions  for  the   fermionic  scattering  states  and  general
expressions  for  the  transmission  and  reflection  coefficients.
In  Sec.~\ref{sec:IV},  we  study  the  fermionic  bound  states:  the  general
properties of the fermionic bound  states  are  described, and their symmetries
under the  charge  conjugation   and   parity  transformations are established.
A condition for  the  existence  of  the  fermionic  zero  modes  is found, and
the properties of these modes are studied.
In Sec.~\ref{sec:V}, we present numerical  results relating to the fermion-kink
interaction.
In the final section, we  briefly  summarize the results obtained in this work.
Appendix A contains some necessary information on the plane-wave states of free
fermions, and  Appendix B  gives  an  explanation  of  the  dependence  of  the
energy of the fermionic bound state on the fermion mass at a qualitative level.

Throughout the paper, the natural units $c = 1$ and $\hbar = 1$ are used.

\section{\label{sec:II}     Lagrangian, symmetries, and  field equations of the
model}

The Lagrangian density  of  the $(1 + 1)$-dimensional sine-Gordon model has the
form
\begin{equation}
\mathcal{L}=\frac{1}{2}\partial _{\mu }\phi \partial ^{\mu }\phi
-m^{4}\lambda^{-1}\bigl(1 - \cos \bigl(m^{-1}\lambda^{1/2}\phi \bigr)
\bigr),                                                            \label{II:1}
\end{equation}
where $\phi(t,x)$ is a real scalar field, $m$  is the mass of the $\phi$-meson,
and $\lambda$ is the  self-interaction  coupling  constant  of  the real scalar
field $\phi$.
The Lagrangian $L = \int \mathcal{L} dx$  possesses  symmetries  under  the two
discrete transformations:
\begin{equation}
\bar{\phi}\left(t,x\right)\rightarrow -\bar{\phi}\left(t,x\right), \label{II:2}
\end{equation}
and
\begin{equation}
\bar{\phi}\left(t,x\right)\rightarrow \bar{\phi} \left(t,x\right) + 2 \pi N,
\; N \in \mathbb{Z},                                               \label{II:3}
\end{equation}
where the rescaled scalar field $\bar{\phi} = m^{-1}\lambda^{1/2}\phi$.
According to these symmetries, the  classical  vacua  of model (\ref{II:1}) are
located at the points
\begin{equation}
\bar{\phi} \left( t,x\right) = 2 \pi N,\; N \in \mathbb{Z},         \label{II:4}
\end{equation}
which correspond to the zero minima of  the  potential $U\left(\bar{\phi}\right)
= 1 - \cos\left(\bar{\phi}\right)$.
It follows from Eqs.~(\ref{II:3}) and (\ref{II:4}) that the field $\bar{\phi} =
m^{-1}\lambda^{1/2}\phi$ may  be interpreted  as  an  angular variable, defined
modulo $2\pi$.
We want the Lagrangian to remain   invariant   under   discrete transformations
(\ref{II:3}) when fermions are included  in  the  sine-Gordon model, and  it is
obvious that the usual version of  the Yukawa interaction $\bar{\psi}\phi \psi$
does not satisfy this requirement.

The interpretation of  $\bar{\phi}$  as  an  angular variable makes it possible
to formulate the sine-Gordon model as  a  nonlinear $\sigma$ model whose target
space is a circle \cite{Manton}.
To do this, we introduce the two-component isovector scalar field
\begin{equation}
\boldsymbol{\phi }=\left( \phi _{1},\phi _{2}\right) =
\bigl( m\lambda^{-1/2}\cos \left( \bar{\phi }\right),\, m\lambda^{-1/2}
\sin \left(\bar{\phi }\right) \bigr),                              \label{II:5}
\end{equation}
which is invariant under discrete transformations (\ref{II:3}).
We want the interaction  between the Dirac fermionic field and isovector scalar
field (\ref{II:5}) to also  be  invariant  under transformations  (\ref{II:3}).
This is most naturally achieved through  the  Yukawa interaction  $\boldsymbol{
\phi} \!\boldsymbol{\cdot}\! \bar{\psi}  \boldsymbol{\tau}_{\perp} \psi$, where
$\boldsymbol{\tau}_{\perp} = (\tau_{1}, \tau_{2})$  and  $\tau_{1, 2}$  are the
corresponding Pauli matrices.
Indeed, the $SO(2)$ isovector $\bar{\psi}\boldsymbol{\tau}_{\perp}\psi$ is also
invariant under the $2 \pi$ rotation about the third  isotopic  axis, as is the
$SO(2)$ isovector $\boldsymbol{\phi}$.
Moreover, the term $\boldsymbol{\phi}\!\boldsymbol{\cdot}\!\bar{\psi}\boldsymbol{
\tau} \psi$ is invariant  under  the full $SU(2)$ isotopic group, and hence the
term $\boldsymbol{\phi}\!\boldsymbol{\cdot}\!\bar{\psi}\boldsymbol{\tau}_{\perp
} \psi$ is invariant under the $U(1)$  subgroup  of  the $SU(2)$ isotopic group
that corresponds to isotopic rotations about the third axis.

In terms of the isovector field $\boldsymbol{\phi }$, the Lagrangian density of
the sine-Gordon model including  an  isodublet  of the Dirac fermions takes the
form
\begin{eqnarray}
\mathcal{L} &=&\frac{1}{2}\partial _{\mu }\boldsymbol{\phi \cdot }
\partial^{\mu}\boldsymbol{\phi} - m^{4}\lambda^{-1}\bigl(1 - m^{-1}
\lambda^{1/2}\phi_{1}\bigr)                                        \label{II:6}
\\
&&+\frac{\nu}{2}\left( \boldsymbol{\phi \cdot \phi }-m^{2}\lambda
^{-1}\right) +i\bar{\psi }\gamma ^{\mu }\partial_{\mu }\psi
- g \boldsymbol{\phi \cdot}\bar{\psi}\boldsymbol{\tau}_{\perp}\psi, \nonumber
\end{eqnarray}
where the Lagrange multiplier  $\nu$  is  introduced to constrain $\boldsymbol{
\phi}$ to lie on the circle $\boldsymbol{\phi \cdot \phi} = m^{2}\lambda^{-1}$.
It is understood that the two types of indices of the Dirac  field $\psi_{i a}$
correspond to its spinor-isospinor structure.
In $(1+1)$ dimensions, we use the following Dirac matrices:
\begin{equation}
\gamma^{0}=\sigma_{1},\; \gamma^{1}=-i\sigma_{2},\;
\gamma_{5}=\gamma^{0}\gamma^{1}=\sigma_{3},                        \label{II:7}
\end{equation}
where $\sigma_{i}$ are the Pauli matrices.
To distinguish the Pauli  matrices  $\sigma_{k}$ acting on the spinor index $i$
of the fermionic field $\psi_{i a}$ from  those  acting  on its isospinor index
$a$, we denote the latter as $\tau_{k}$.

It is readily seen that the Lagrangian  density (\ref{II:6}) is invariant under
discrete $\mathbb{Z}$-transformations (\ref{II:3}).
It is also invariant under discrete $\mathbb{Z}_{2}$-transformation (\ref{II:2})
provided that the Dirac fermionic field  $\psi_{i a}$  is  transformed as $\psi
\rightarrow \mathbb{I}\otimes \tau_{3}\psi$.
By varying the action $S = \int \mathcal{L} dx dt$ in the corresponding fields,
we obtain the field equations:
\begin{equation}
\partial_{\mu}\partial^{\mu}\boldsymbol{\phi} - m^{3}\lambda^{-1/2}
\mathbf{n}_{1}- \nu \boldsymbol{\phi} + g \bar{\psi}
\boldsymbol{\tau}_{\perp}\psi = 0,                                 \label{II:8}
\end{equation}
and
\begin{equation}
i\gamma^{\mu }\partial_{\mu }\psi - g \boldsymbol{\phi }\boldsymbol{\cdot}
\boldsymbol{\tau}_{\perp }\psi = 0,                                \label{II:9}
\end{equation}
where the Lagrange multiplier
\begin{equation}
\nu =  m^{-2}\lambda \bigl(\boldsymbol{\phi}
\!\boldsymbol{\cdot}\!\partial_{\mu }\partial^{\mu}\boldsymbol{\phi}-
m^{3}\lambda^{-1/2}\boldsymbol{\phi}
\!\boldsymbol{\cdot}\!\mathbf{n}_{1}
 + g \boldsymbol{\phi}\!\boldsymbol{\cdot}\! \bar{\psi}
\boldsymbol{\tau}_{\perp}\psi \bigr)                              \label{II:10}
\end{equation}
and the unit isovector $\mathbf{n}_{1} = (1, 0)$.

It is known that that among  the  numerous  soliton  solutions, the sine-Gordon
model possesses the static kink solution
\begin{equation}
\phi _{\text{k}}(x) = 4m\lambda^{-1/2}\arctan\left[\exp\left(m x \right)
\right]                                                          \label{II:11}
\end{equation}
or in terms of the isovector field
\begin{eqnarray}
\boldsymbol{\phi}_{\text{k}}(x) & = & m\lambda^{-1/2}
\left(1 - 2\text{sech}^{2}\left( m x\right) ,\right. \nonumber \\
&&\left. -2\text{sech}\left( m x\right)
\tanh \left( m x\right) \right).                                  \label{II:12}
\end{eqnarray}
The sine-Gordon kink has mass
\begin{equation}
M_{\text{k}} = 8 m^{3} \lambda^{-1}.                              \label{II:13}
\end{equation}
Note that the mass of the kink is inversely proportional to the self-interaction
coupling constant, which is a  characteristic  property  of  soliton  solutions
\cite{Rajaraman}.

We shall  now  establish  the  symmetry   properties   of   the  Dirac equation
(\ref{II:9}) under the discrete $C$, $P$, and  $T$  transformations in the case
of external field of kink solution (\ref{II:12}).
It is easy to see that the kink solution has the symmetry property
\begin{equation}
\boldsymbol{\phi }_{\text{k}}\left( x\right)
\!\boldsymbol{\cdot}\!\boldsymbol{\tau}_{\perp }=
\boldsymbol{\phi}_{\text{k}}\left(-x\right)
\!\boldsymbol{\cdot}\!\boldsymbol{\tau }_{\perp}^{\ast}.          \label{II:14}
\end{equation}
Using this symmetry property, it can easily be shown  that if $\psi(t, x)$ is a
solution  of  the  Dirac  equation (\ref{II:9})  in  the external field of kink
(\ref{II:12}), then
\begin{subequations}                                              \label{II:15}
\begin{eqnarray}
\psi ^{C}\left( t,x\right)  &=&\eta _{C}\gamma _{5}\otimes \tau _{1}\psi
^{\ast }\left( t,x\right),                                       \label{II:15a}
\\
\psi ^{P}\left( t,x\right)  &=&\eta _{P}\gamma ^{0}\otimes \tau _{1}\psi
\left( t,-x\right),                                              \label{II:15b}
\\
\psi ^{T}\left( t,x\right)  &=&\eta _{T}\gamma ^{0}\otimes \tau _{1}\psi
^{\ast }\left(-t,x\right),                                      \label{II:15c}
\end{eqnarray}
\end{subequations}
where $\eta_{C}$, $\eta_{P}$, and $\eta_{T}$  are  phase  multipliers, are also
solutions of the Dirac equation in the external field of the kink.
Of course, the Lagrangian  density (\ref{II:6}) remains  invariant  under these
$C$, $P$, and $T$ transformations.

The antikink solution is obtained from  kink solution (\ref{II:12}) through the
inversion $x \rightarrow -x$.
Using the symmetry property  (\ref{II:14}),  it  can  easily  be  shown that if
$\psi_{\text{k}}(t, x)$ is a solution of the Dirac equation (\ref{II:9}) in the
external field of kink (\ref{II:12}), then
\begin{equation}
\psi _{\text{ak}}\left(t, x\right) =\mathbb{I}\otimes \tau_{1}
\psi_{\text{k}}\left(t, x\right)                                 \label{II:15d}
\end{equation}
is a solution of the Dirac equation (\ref{II:9}) in the external  field  of the
antikink $\boldsymbol{\phi}_{\text{ak}}(x) = \boldsymbol{\phi}_{\text{k}}(-x)$.

As $x \rightarrow \pm \infty$, isovector  kink field (\ref{II:12}) tends to the
vacuum value
\begin{equation}
\underset{x\rightarrow \pm \infty }{\lim}\boldsymbol{\phi}_{\text{k}}
\left(x\right)=\boldsymbol{\phi}_{\text{vac}} =
\bigl(m\lambda^{-1/2},0\bigr).                                    \label{II:16}
\end{equation}
It follows that in  the  distant  spatial regions $\left\vert x \right\vert \gg
m^{-1}$, the Dirac equation (\ref{II:9}) describes free fermions (antifermions)
with mass $M = m g \lambda^{-1/2}$.
The properties of the corresponding plane-wave solutions  of the Dirac equation
and their explicit forms are given in Appendix A.

\section{\label{sec:III} Fermion-kink scattering}

We shall consider fermion  scattering  on the sine-Gordon kink (\ref{II:12}) in
the external field approximation.
In this approximation, the backreaction of  a  fermion on the field of the kink
is neglected, and  kink-soliton  scattering  is  described  solely by the Dirac
equation (\ref{II:9}).
To neglect the fermion backreaction, the contribution of the fermionic terms in
Eqs.~(\ref{II:8}) and (\ref{II:10}) should  be  much  less  than  that  of  the
corresponding bosonic terms.
Using normalization condition (\ref{A:4c}) to  estimate the contribution of the
fermionic terms, it can be shown that  the condition of their smallness has the
form
\begin{equation}
g^{2} \ll m^{2} \varepsilon L,                                    \label{III:I}
\end{equation}
where $\varepsilon$ is the fermion  energy  and  $L$  is the normalized length.
In the nonrelativistic case,  condition (\ref{III:I})  becomes  more stringent:
\begin{equation}
g^{2} \ll m^{2} M L = g m^{2}\lambda^{-1/2} m L,                 \label{III:II}
\end{equation}
where the fermion mass $M = m g \lambda^{-1/2}$.

We see that conditions (\ref{III:I}) and (\ref{III:II}) can always be fulfilled
if the normalized  length  $L$  is  sufficiently  large, corresponding to a low
linear density of the incident fermions.
From the viewpoint of QFT, however, we are  talking  about  the scattering of a
fermion  of  mass $M  =  m g \lambda^{-1/2}$  on  a  sine-Gordon  kink  of mass
$M_{\text{k}} = 8 m^{3} \lambda^{-1}$.
To neglect the recoil of  the  kink in fermion scattering, the mass of the kink
$M_{\text{k}}$  must be  much  larger  than  the mass of the fermion $M$, which
leads to the condition
\begin{equation}
g \ll 8 m^{2} \lambda^{-1/2}.                                   \label{III:III}
\end{equation}
By comparing Eqs.~(\ref{III:II})  and  (\ref{III:III}),  we  see  that they are
equivalent if the normalization length $L \approx 8 m^{-1}$.
In turn, it follows from Eq.~(\ref{II:11}) that the value of $8 m^{-1}$  can be
interpreted as the spatial size of the kink.
Hence, Eqs.~(\ref{III:II}) and (\ref{III:III}) become equivalent  when there is
on average one fermion in a region whose length is equal to the spatial size of
the kink.

Substituting the ansatz $\psi_{i a}\left(t, x \right) = \exp\left(-i\varepsilon
t \right)\psi_{i a}\left( x \right)$ and  background  kink  field (\ref{II:12})
into the Dirac equation (\ref{II:9}), we obtain  the  two  independent  systems
of differential equations:
\begin{equation}
\begin{pmatrix}
\psi _{11}^{\prime } \\
\psi _{22}^{\prime }
\end{pmatrix}
=
\begin{pmatrix}
i\varepsilon  & -i M F\left( x\right)  \\
i M F^{\ast}\!\!\left( x\right) & -i\varepsilon
\end{pmatrix}%
\begin{pmatrix}
\psi_{11} \\
\psi_{22}
\end{pmatrix}                                                     \label{III:1}
\end{equation}
and
\begin{equation}
\begin{pmatrix}
\psi _{12}^{\prime } \\
\psi _{21}^{\prime }
\end{pmatrix}
=
\begin{pmatrix}
i\varepsilon  & -i M F^{\ast}\!\!\left( x\right) \\
i M F\left( x\right)  & -i\varepsilon
\end{pmatrix}
\begin{pmatrix}
\psi _{12} \\
\psi _{21}
\end{pmatrix},                                                    \label{III:2}
\end{equation}
where the unitary function
\begin{eqnarray}
F(x) &=&\left( \frac{e^{mx}+i}{e^{mx}-i}\right)^{2}               \label{III:3}
\\
&=&1 - 2\,\text{sech}^{2}\left(m x\right) + 2 i\,\text{sech}(mx) \tanh(mx).
                                                                  \nonumber
\end{eqnarray}
The  Dirac  equation  (\ref{II:9})  can  be  split  into  the  two  independent
subsystems (\ref{III:1}) and (\ref{III:2}) since the Dirac Hamiltonian
\begin{equation}
H_{D}=\alpha \!\otimes\! \mathbb{I}\left(-i\partial_{x}\right) +
g \beta \!\otimes\! \boldsymbol{\phi}_{\text{k}}\!\boldsymbol{\cdot}\!
\boldsymbol{\tau}_{\perp}                                         \label{III:4}
\end{equation}
commutes with  the spin-isospin operator $T_{3}=\gamma_{5}\!\otimes\!\tau_{3}$:
\begin{equation}
\left[H_{D}, T_{3}\right] = 0.                                    \label{III:5}
\end{equation}
Here, Eq.~(\ref{III:1})  contains  only  the  diagonal  elements  of the matrix
$\psi_{i a}$, whereas Eq.~(\ref{III:2}) contains  only  the  antidiagonal ones.
Let us denote the diagonal and antidiagonal parts of the matrix $\psi_{i a}$ as
$\psi_{\text{d}}$ and $\psi_{\text{a}}$, respectively:
\begin{equation}
\psi_{\text{d}}=
\begin{pmatrix}
\psi _{11} & 0 \\
0 & \psi _{22}
\end{pmatrix}
,\quad \psi_{\text{a}}=
\begin{pmatrix}
0 & \psi _{12} \\
\psi _{21} & 0
\end{pmatrix}.                                                    \label{III:6}
\end{equation}
It can then be easily   shown  that $\psi_{\text{d}}$ and $\psi_{\text{a}}$ are
the eigenmatrices of the operator $T_{3}$:
\begin{equation}
T_{3}\psi_{\text{d}}=\psi_{\text{d}}, \quad
T_{3}\psi_{\text{a}}=-\psi_{\text{a}}.                            \label{III:7}
\end{equation}
It follows from Eqs.~(\ref{III:5}) and (\ref{III:7}) that $\psi_{\text{d}}$ and
$\psi_{\text{a}}$ are  independent  from  each  other, and this is reflected in
Eqs.~(\ref{III:1}) and (\ref{III:2}).

It can be shown that system (\ref{III:1})  is  equivalent  to  the second order
differential equation
\begin{eqnarray}
&\psi _{11}^{\prime \prime }(x)+2 i m \, \text{sech}(m x)\psi_{11}^{\prime}(x)&
\nonumber \\
&+\left(k^{2}+2\varepsilon m \,\text{sech}(m x)\right)
\psi_{11}(x)& = 0                                                 \label{III:8}
\end{eqnarray}
together with the relation
\begin{equation}
\psi_{22}(x)=M^{-1}F^{\ast}\!\!\left(x\right)\left(\varepsilon \psi
_{11}(x) + i \psi_{11}^{\prime }(x)\right),                       \label{III:9}
\end{equation}
where $k^{2} = \varepsilon^{2} - M^{2}$.
Specifically, any solution to system (\ref{III:1}) satisfies Eqs.~(\ref{III:8})
and (\ref{III:9}).
Conversely,  the  function   $\psi_{11}(x)$  that  satisfies  Eq.~(\ref{III:8})
together with the  function  $\psi_{22}(x)$  obtained   from  Eq.~(\ref{III:9})
form a solution to system (\ref{III:1}).
Indeed, in the latter case, the derivative  $\psi_{22}^{\prime}$ satisfies the
relation
\begin{equation}
\psi _{22}^{\prime }\left( x\right) =M^{-1}F^{\ast }\left( x\right) \left(
\varepsilon \psi _{11}^{\prime }\left( x\right) -ik^{2}\psi _{11}\left(
x\right) \right),                                                \label{III:9a}
\end{equation}
where  we  use  Eq.~(\ref{III:8})  to   eliminate  $\psi_{11}^{\prime \prime}$.
It is easy to see that  by  substituting  Eqs.~(\ref{III:9}) and (\ref{III:9a})
into system (\ref{III:1}), the latter is identically satisfied.
Similarly, system (\ref{III:2}) is equivalent  to the second-order differential
equation
\begin{eqnarray}
&\psi_{12}^{\prime \prime}(x) - 2 i m \,\text{sech}(mx)\psi _{12}^{\prime }(x)&
\nonumber \\
& + \left(k^{2} - 2 \varepsilon m\,
\text{sech}(mx)\right)\psi_{12}(x)& = 0                          \label{III:10}
\end{eqnarray}
together with the relation
\begin{equation}
\psi _{21}(x)= M^{-1}F\left( x\right) \left( \varepsilon \psi _{12}(x)+i\psi
_{12}^{\prime }(x)\right).                                       \label{III:11}
\end{equation}

By applying a change of variable $\xi=\tanh\left( m x\right)$, the second-order
differential equations  (\ref{III:8})  and  (\ref{III:10})  are  reduced to the
forms
\begin{gather}
\psi _{11}^{\prime \prime }(\xi )-\frac{2\left( \xi -i\sqrt{1-\xi ^{2}}
\right) }{1-\xi ^{2}}\psi _{11}^{\prime }(\xi ) \nonumber \\
+\frac{\left( k^{2}+2\varepsilon m\sqrt{1-\xi ^{2}}\right) }{m^{2}\left(
1-\xi ^{2}\right) ^{2}}\psi _{11}(\xi )=0                        \label{III:12}
\end{gather}
and
\begin{gather}
\psi _{12}^{\prime \prime }(\xi )-\frac{2\left(\xi + i \sqrt{1 - \xi^{2}}
\right) }{1-\xi ^{2}}\psi _{12}^{\prime }(\xi ) \nonumber \\
+\frac{\left( k^{2}-2\varepsilon m\sqrt{1-\xi ^{2}}\right)}{m^{2}\left(
1-\xi ^{2}\right) ^{2}}\psi _{12}(\xi) = 0.                      \label{III:13}
\end{gather}
The solutions to  differential  equations (\ref{III:12}) and (\ref{III:13}) can
be expressed in terms of the local Heun functions \cite{Ronveaux, DLMF}.
Let the fermionic wave fall on the kink from the left.
Then, as $x \rightarrow + \infty$,  the  solutions  to  Eqs.~(\ref{III:12}) and
(\ref{III:13}) must approximate the transmitted plane  wave,  which is $\propto
\exp(ikx)$.
In  terms  of  the  initial   variable  $x$,  the  corresponding  solutions  to
Eqs.~(\ref{III:12}) and  (\ref{III:13}) are 
\begin{widetext}
\begin{equation}
\psi_{11}\left( x \right) =e^{ikx}Hl\left[ \frac{1}{2},-i\frac{2\left(
\varepsilon -k\right) }{m};-1,0,1-i\frac{2k}{m},1+i\frac{2k}{m};\frac{1}{
1+ie^{mx}}\right]                                                \label{III:14}
\end{equation}
and
\begin{eqnarray}
\psi_{12}\left( x \right) &=&e^{-\frac{\pi }{2}\frac{k}{m}}
\left(e^{-i\frac{\pi}{4}}+e^{i\frac{\pi}{4}}e^{-mx}\right)^{2i\frac{k}{m}}
e^{ikx}\times \nonumber
\\
&&Hl\left[ \frac{1}{2},-2\frac{k^{2}}{m^{2}}-i\frac{2\varepsilon -k}{m};-1-i
\frac{2k}{m},-i\frac{2k}{m},1-i\frac{2k}{m},1-i\frac{2k}{m};\frac{1}
{1 - i e^{mx}}\right],                                           \label{III:15}
\end{eqnarray}
\end{widetext}
where we use the notation $Hl\left(a, q; \alpha, \beta,\gamma,\delta ;z\right)$
for the six-parameter local Heun function \cite{Ronveaux, DLMF}.

The local  Heun function $Hl\left(a, q; \alpha, \beta,\gamma,\delta; z \right)$
represents the  solution  to  the  second-order  differential  Heun's  equation
\cite{Ronveaux, DLMF}.
Heun's equation  possesses  four  regular  singular  points located at $z = 0$,
$z = 1$, $z = a$, and $z = \infty$.
In Eqs.~(\ref{III:14}) and  (\ref{III:15}),  the  arguments  of  the local Heun
functions tend to zero (one of the regular  singular  points) as $x \rightarrow
+\infty$.
The local Heun function $Hl\left(a, q; \alpha, \beta, \gamma, \delta; z\right)$
is analytic, and is equal to one at the regular singular point $z =0$, where it
can be expanded in a Taylor series.
In the complex  $z$-plane, the radius of convergence of this series is equal to
the smallest of $\left\vert a \right\vert$ and one.
It follows from  Eqs.~(\ref{III:14}) and (\ref{III:15}) that in our case, it is
equal to $1/2$.

Despite the finite radius of convergence of the series, the local Heun function
$Hl\left(1/2, q; \alpha, \beta, \gamma, \delta; z\right)$  can  be analytically
continued to the  whole complex  plane with the cut $\left[ 1/2,\infty\right)$.
It will be defined at all finite points  of  the  complex  plane except for the
regular singular points $z = 1/2$ and $z = 1$.
At the same time, in Eqs.~(\ref{III:14})  and  (\ref{III:15}), the arguments of
the local Heun functions  tend  to  one  as  $x \rightarrow -\infty$, and hence
local  solutions  (\ref{III:14})  and  (\ref{III:15})  cannot  be  used  as  $x
\rightarrow -\infty$.

It is known \cite{Ronveaux,  DLMF},  however, that Heun's equation  has another
local solution such that the local Heun functions  entering it will be analytic
in the neighborhood of the singular point $z = 1$. 
The same  is   also   true   for   the   solutions  to  Eqs.~(\ref{III:12}) and
(\ref{III:13}).
Specifically, it  can  be shown that  the solutions $\psi_{11}$ and $\psi_{12}$
can be represented in the alternative forms
\begin{widetext}
\begin{eqnarray}
\psi_{11} & = & c_{1}e^{ikx}Hl\left[ \frac{1}{2},i\frac{%
2\left( \varepsilon -k\right) }{m};-1,0,1+i\frac{2k}{m},1-i\frac{2k}{m};
\frac{1}{1-ie^{-mx}}\right] + \nonumber \\
&&c_{2}e^{-ikx}\left( -i+e^{mx}\right)^{2i\frac{k}{m}}Hl\left[ \frac{1}{2}
,-2\frac{k^{2}}{m^{2}}+i\frac{2\varepsilon +k}{m};-1-i\frac{2k}{m},-i\frac{2k
}{m},1-i\frac{2k}{m},1-i\frac{2k}{m};\frac{1}{1-ie^{-mx}}\right] \label{III:16}
\end{eqnarray}
and
\begin{eqnarray}
\psi _{12} &=&d_{1}e^{-\pi \frac{k}{m}}e^{ikx}\left( i+e^{mx}\right) ^{-2i%
\frac{k}{m}}Hl\left[ \frac{1}{2},-2\frac{k^{2}}{m^{2}}+i\frac{2\varepsilon -k
}{m};-1+i\frac{2k}{m},i\frac{2k}{m},1+i\frac{2k}{m},1+i\frac{2k}{m};\frac{1}{
1+ie^{-mx}}\right] + \nonumber \\
&&d_{2}e^{-\pi \frac{k}{m}}e^{-ikx}Hl\left[ \frac{1}{2},i\frac{2\left(
k+\varepsilon \right) }{m};-1,0,1-i\frac{2k}{m},1+i\frac{2k}{m};\frac{1}
{1+ie^{-mx}}\right],                                             \label{III:17}
\end{eqnarray}
\end{widetext}
where  $c_{1}$,  $c_{2}$,  $d_{1}$,  and  $d_{2}$  are  constant  coefficients.
Now the arguments of the local Heun  functions  tend  to zero as $x \rightarrow
-\infty$, meaning that the local Heun functions remain  analytic in this limit.
Forms (\ref{III:14})  and  (\ref{III:16})  of  the  solution  $\psi_{11}$  have
overlapping domains  of  analyticity  in  the  variable  $x$.
The same  is  also  true  for  forms  (\ref{III:15})  and (\ref{III:17}) of the
solution $\psi_{12}$.  
It follows that by equating the  alternative expressions for $\psi_{11}(x)$ and
its derivative $d\psi_{11}(x)/dx$, we  can  find  the  coefficients $c_{1}$ and
$c_{2}$ in Eq.~(\ref{III:16}), and in Eq.~(\ref{III:17}), the coefficients $d_{
1}$ and $d_{2}$  can be found in a similar way.
The $x$-coordinate of the matching  point  does  not matter, and can lie in the
interval $(-\infty, \infty)$.
For reasons of symmetry, we choose  the  coordinate  of  the  matching point as
$x = 0$.
In this case, the coefficients $c_{1}$ and $c_{2}$ can be written as
\begin{widetext}
\begin{eqnarray}
c_{1} &=&\frac{4sk\chi _{\text{tr}}\left( s^{\ast }\right) \chi _{
\text{rf}}\left( s\right) -m\left[ \chi _{\text{tr}}^{\prime }\left( s^{\ast
}\right) \chi _{\text{rf}}\left( s\right) +\chi _{\text{tr}}\left( s^{\ast
}\right) \chi _{\text{rf}}^{\prime }\left( s\right) \right] }{4sk\chi _{
\text{rf}}\left( s\right) \chi _{\text{in}}\left( s\right) + m
W[\chi_{\text{rf}}\left(s\right),\chi_{\text{in}}\left(s\right)]},
                                                                 \label{III:18}
\\
c_{2} &=&2^{-i\frac{k}{m}}e^{-\frac{\pi }{2}\frac{k}{m}}\frac{m\left[
\chi _{\text{tr}}^{\prime }\left( s^{\ast }\right) \chi _{\text{in}}\left(
s\right) +\chi _{\text{tr}}\left( s^{\ast }\right) \chi _{\text{in}}^{\prime
}\left( s\right) \right] }{4sk\chi _{\text{rf}}\left( s\right) \chi _{\text{
in}}\left( s\right) + m W[\chi_{\text{rf}}\left( s\right) ,\chi_{\text{in}
}\left( s\right) ]},                                             \label{III:19}
\end{eqnarray}
where the functions
\begin{eqnarray}
\chi _{\text{in}}\left( z\right)  &=&Hl\left[ \frac{1}{2},i\frac{2\left(
\varepsilon -k\right) }{m};-1,0,1+i\frac{2k}{m},1-i\frac{2k}{m};z\right],
                                                                 \label{III:20}
\\
\chi _{\text{rf}}\left( z\right)  &=&Hl\left[ \frac{1}{2},-2\frac{k^{2}}{
m^{2}}+i\frac{2\varepsilon +k}{m};-1-i\frac{2k}{m},-i\frac{2k}{m},
1-i\frac{2k}{m},1-i\frac{2k}{m};z\right],                        \label{III:21}
\\
\chi _{\text{tr}}\left( z\right)  &=&Hl\left[ \frac{1}{2},-i\frac{2\left(
\varepsilon -k\right) }{m};-1,0,1-i\frac{2k}{m},
1+i\frac{2k}{m};z\right],                                        \label{III:22}
\end{eqnarray}
the Wronskian
\begin{equation}
W[\chi _{\text{rf}}\left( s\right) ,\chi _{\text{in}}\left( s\right) ]=\left[
\chi _{\text{rf}}\left( z\right) \chi _{\text{in}}^{\prime }\left( z\right)
-\chi _{\text{rf}}^{\prime }\left( z\right) \chi _{\text{in}}\left( z\right)
\right] _{z=s},                                                  \label{III:23}
\end{equation}
\end{widetext}
and the variable $s = \left(1 + i\right)/2$.
Similarly, the coefficients $d_{1}$ and $d_{2}$ are
\begin{widetext}
\begin{eqnarray}
d_{1} &=&-2^{2i\frac{k}{m}}\frac{4sk\eta _{\text{tr}}\left( s\right) \eta _{
\text{rf}}\left( s^{\ast }\right) -m\left[ \eta _{\text{tr}}^{\prime }\left(
s\right) \eta _{\text{rf}}\left( s^{\ast }\right) +\eta _{\text{tr}}\left(
s\right) \eta _{\text{rf}}^{\prime }\left( s^{\ast }\right) \right] }{
4s^{\ast }k\eta _{\text{rf}}\left( s^{\ast }\right) \eta _{\text{in}}\left(
s^{\ast }\right) -mW[\eta _{\text{rf}}\left( s^{\ast }\right) ,\eta _{\text{
in}}\left( s^{\ast }\right) ]},                                  \label{III:24}
\\
d_{2} &=&2^{i\frac{k}{m}}e^{\frac{\pi }{2}\frac{k}{m}}\frac{4k\eta _{\text{tr
}}\left( s\right) \eta _{\text{in}}\left( s^{\ast }\right) -m\left[ \eta _{
\text{tr}}^{\prime }\left( s\right) \eta _{\text{in}}\left( s^{\ast }\right)
+\eta _{\text{tr}}\left( s\right) \eta _{\text{in}}^{\prime }\left( s^{\ast
}\right) \right] }{4s^{\ast }k\eta _{\text{rf}}\left( s^{\ast }\right) \eta
_{\text{in}}\left( s^{\ast }\right) - m W[\eta_{\text{rf}}\left(s^{\ast
}\right) ,\eta _{\text{in}}\left( s^{\ast}\right)]},             \label{III:25}
\end{eqnarray}
where the functions
\begin{eqnarray}
\eta _{\text{in}}\left( z\right)  &=&Hl\left[ \frac{1}{2},-2\frac{k^{2}}{
m^{2}}+i\frac{2\varepsilon -k}{m};-1+i\frac{2k}{m},i\frac{2k}{m},1+i\frac{2k
}{m},1+i\frac{2k}{m};z\right],                                   \label{III:26}
\\
\eta _{\text{rf}}\left( z\right)  &=&Hl\left[ \frac{1}{2},i\frac{2\left(
k+\varepsilon \right) }{m};-1,0,1-i\frac{2k}{m},1+i\frac{2k}{m};z\right],
                                                                \label{III:26a}
\\
\eta _{\text{tr}}\left( z\right)  &=&Hl\left[ \frac{1}{2},-2\frac{k^{2}}{
m^{2}}-i\frac{2\varepsilon -k}{m};-1-i\frac{2k}{m},-i\frac{2k}{m},1-i\frac{2k
}{m},1-i\frac{2k}{m};z\right],                                   \label{III:27}
\end{eqnarray}
the Wronskian
\begin{equation}
W[\eta _{\text{rf}}\left( s^{\ast }\right) ,\eta _{\text{in}}\left( s^{\ast
}\right) ]=\left[ \eta _{\text{rf}}\left( z\right) \eta _{\text{in}}^{\prime
}\left( z\right) -\eta _{\text{rf}}^{\prime }\left( z\right) \eta _{\text{in}
}\left( z\right) \right] _{z=s^{\ast }},                         \label{III:28}
\end{equation}
\end{widetext}
and the variable $s^{\ast} = \left(1 - i\right)/2$.
In  Eqs.~(\ref{III:18})--(\ref{III:28}),  the  subscripts  ``in'',  ``rf'', and
``tr''  refer  to  the  incident,  transmitted,  and reflected fermionic  wave,
respectively.

The coefficients  $c_{1}$  and  $c_{2}$  ($d_{1}$ and $d_{2}$) contain  all the
information about  the  scattering  of  the  diagonal  (antidiagonal) component
$\psi_{\text{d}}$ ($\psi_{\text{a}}$) of  the fermionic wave on the sine-Gordon
kink.
It follows from Eqs.~(\ref{III:14})  and  (\ref{III:16})  that  the asymptotics
of scattering for the  $\psi_{11}$ component  can  be schematically represented
as
\begin{equation}
c_{1}e^{ikx} \rightarrow e^{ikx} + c_{2}e^{\pi \frac{k}{m}}e^{-i k x},
                                                                 \label{III:29}
\end{equation}
which corresponds to the splitting  of  the  incident wave into the transmitted
and reflected waves.
Next, using Eq.~(\ref{III:9})  for the $\psi_{22}$ component and the expression
$j^{\mu} = \bar{\psi}\gamma^{\mu}\!\otimes\!\mathbb{I} \psi = \left(\psi^{\ast}
\psi, \psi^{\ast}\sigma_{3}\!\otimes\!\mathbb{I}\psi \right)$ for the fermionic
current, we obtain  expressions  for  the  incident, transmitted, and reflected
currents of the diagonal component $\psi_{\text{d}}$:
\begin{eqnarray}
j_{\text{in}} &=&c_{1}c_{1}^{\ast }\bigl( 1-M^{-2}\left( \varepsilon
-k\right) ^{2}\bigr),                                            \label{III:30}
\\
j_{\text{rf}} &=&c_{2}c_{2}^{\ast }e^{2\pi \frac{k}{m}}\bigl(1 - M^{-2}\left(
\varepsilon + k \right)^{2} \bigr),                              \label{III:31}
\\
j_{\text{tr}} &=&1-M^{-2}\left(\varepsilon - k\right)^{2}.       \label{III:32}
\end{eqnarray}
Using  Eqs.~(\ref{III:30})--(\ref{III:31}),  we  obtain   expressions  for  the
transmission and reflection coefficients:
\begin{eqnarray}
T &=&\frac{j_{\text{tr}}}{j_{\text{in}}}=\left\vert c_{1}\right\vert^{-2},
                                                                 \label{III:33}
\\
R &=&\frac{\left\vert j_{\text{rf}}\right \vert}{j_{\text{in}}} =
\frac{\left\vert c_{2}\right\vert^{2}}{\left\vert c_{1}\right\vert ^{2}}
e^{2\pi \frac{k}{m}}\left(1+2M^{-2}k\left( \varepsilon +k\right) \right),
                                                                 \label{III:34}
\end{eqnarray}
which correspond to the scattering of  the diagonal component $\psi_{\text{d}}$
on the sine-Gordon kink.
In the process of scattering, the transmitted diagonal fermionic  wave acquires
a  phase  shift  $\delta_{\text{d}}$  with  respect  to  the  incident diagonal
fermionic wave.
Eq.~(\ref{III:29}) tells us that this phase shift
\begin{equation}
\delta_{\text{d}}(k) = - \text{arg}[c_{1}(k)].                  \label{III:34a}
\end{equation}

In a  similar  way  to  Eq.~(\ref{III:29}),  the  scattering of the $\psi_{12}$
component can be schematically written as
\begin{equation}
d_{1}e^{ikx}  \rightarrow  e^{ikx} + d_{2}e^{-\pi \frac{k}{m}}e^{-ikx}.
                                                                 \label{III:35}
\end{equation}
In the same  way  as  above,  we  can  sequentially  obtain expressions for the
fermionic currents:
\begin{eqnarray}
j_{\text{in}} &=&d_{1}d_{1}^{\ast }\bigl( 1-M^{-2}\left( \varepsilon
-k\right) ^{2}\bigr),                                            \label{III:36}
\\
j_{\text{rf}} &=&d_{2}d_{2}^{\ast }e^{-2\pi \frac{k}{m}}\bigl(
1-M^{-2}\left( \varepsilon +k\right) ^{2} \bigr),                \label{III:37}
\\
j_{\text{tr}} &=&1-M^{-2}\left( \varepsilon -k\right)^{2},       \label{III:38}
\end{eqnarray}
and the corresponding transmission and reflection coefficients:
\begin{eqnarray}
T &=&\frac{j_{\text{tr}}}{j_{\text{in}}}=\left\vert d_{1}\right\vert^{-2},
                                                                 \label{III:39}
\\
R &=&\frac{\left\vert j_{\text{rf}}\right\vert}{j_{\text{in}}}=
\frac{\left\vert d_{2}\right\vert
^{2}}{\left\vert d_{1}\right\vert ^{2}}e^{-2\pi \frac{k}{m}}\left(
1+2M^{-2}k\left( \varepsilon +k\right) \right).                  \label{III:40}
\end{eqnarray}
Eqs.~(\ref{III:36})--(\ref{III:40})  correspond  to   the   scattering  of  the
antidiagonal component $\psi_{\text{a}}$ on the sine-Gordon kink.
From  Eq.~(\ref{III:35}),  it  follows  that  when  the  antidiagonal component
$\psi_{\text{a}}$ is scattered, the phase shift
\begin{equation}
\delta_{\text{a}}(k) = - \text{arg}[d_{1}(k)].                  \label{III:40a}
\end{equation}

Let us now  define  the  mean  value  of  the  isospin  $x$-projection  $I_{1}$
for a plane-wave fermionic state as
\begin{equation}
\left\langle I_{1}\right\rangle = \frac{1}{2}\frac{\psi ^{\ast}\mathbb{I}
\otimes \tau_{1}\psi}{\psi^{\ast}\mathbb{I}\otimes\mathbb{I}\psi}.
                                                                 \label{III:41}
\end{equation}
It follows from  Eq.~(\ref{III:41}) that  the  mean  value  $\left\langle I_{1}
\right\rangle$ of the isospin $x$-projection lies in the interval $[-1/2,1/2]$,
and is relativistically invariant, as it should be.
It can easily be shown that the  mean  value  $\left\langle I_{1}\right\rangle$
vanishes identically  for  the  both  the  diagonal  and antidiagonal fermionic
states:
\begin{eqnarray}
\left\langle I_{1}\right\rangle _{\text{d, in}} &=&\left\langle
I_{1}\right\rangle _{\text{d, rf}}=\left\langle I_{1}\right\rangle_{\text{
d, tr}}=0,                                                       \label{III:42}
\\
\left\langle I_{1}\right\rangle _{\text{a, in}} &=&\left\langle
I_{1}\right\rangle _{\text{a, rf}}=\left\langle I_{1}\right\rangle _{\text{
a, tr}}=0.                                                       \label{III:43}
\end{eqnarray}
However, for a linear combination $\alpha\psi_{\text{d}}+\beta\psi_{\text{a}}$,
the mean  value  $\left\langle  I_{1} \right\rangle$  is  different  from zero.
In particular, it can be shown that
\begin{eqnarray}
\left\langle I_{1}\right\rangle _{\text{in}} &=&\frac{\text{Re}\left[ \alpha
^{\ast }\beta \right] }{\left\vert \alpha \right\vert ^{2}+\left\vert \beta
\right\vert ^{2}},                                               \label{III:44}
\\
\left\langle I_{1}\right\rangle _{\text{tr}} &=&\frac{\text{Re}\left[ \alpha
^{\ast }\beta c_{1}d_{1}^{\ast }\right] }{\left\vert \alpha d_{1}\right\vert
^{2}+\left\vert \beta c_{1}\right\vert ^{2}},                    \label{III:45}
\\
\left\langle I_{1}\right\rangle _{\text{rf}} &=&\frac{\text{Re}\left[ \alpha
^{\ast }\beta c_{1}c_{2}^{\ast }d_{1}^{\ast }d_{2}\right] }{e^{-2\pi \frac{k
}{m}}\left\vert \alpha c_{2}d_{1}\right\vert ^{2}+e^{2\pi \frac{k}{m}
}\left\vert \beta c_{1}d_{2}\right\vert ^{2}}                    \label{III:46}
\end{eqnarray}
for the incident, reflected,  and  transmitted  waves  of  the scattering state
$\alpha \psi_{\text{d}} + \beta \psi_{\text{a}}$, respectively.

\section{\label{sec:IV} Fermionic bound states}

Let us investigate the presence of fermionic bound states in the external field
of the sine-Gordon kink.
It is obvious that the energy $\varepsilon$ of  a fermionic bound state must be
less than the fermion mass $M$.
In this case, the parameter $k^{2} = \varepsilon^{2} - M^{2}$ becomes negative:
$k^{2} \equiv -\kappa^{2} < 0$.
It  then  follows  from  Eqs.~(\ref{III:8}), (\ref{III:9}), (\ref{III:10}), and
(\ref{III:11}) that the components  of  the  wave function of a fermionic bound
state  are  $\propto \exp \left( - \kappa \left\vert x \right\vert \right)$  as
$\left\vert x \right\vert \rightarrow \infty$.
Next, it is easily shown  that  under  charge  conjugation  (\ref{II:15a}), the
diagonal fermionic wave function  becomes  an antidiagonal one, and vice versa.
Since the charge conjugation reverses the sign  of  the energy of the fermionic
state, the bound states of  the  Dirac  Hamiltonian  (\ref{III:4}) can be split
into pairs, each  consisting  of  diagonal  and  antidiagonal bound states with
opposite  energies  and  connected  to   each   other   by  charge  conjugation
(\ref{II:15a}).

Further, it can be  shown  that  the  Dirac  Hamiltonian  (\ref{III:4}), parity
operator (\ref{II:15b}), and operator $T_{3}$ (\ref{III:5}) commute  with  each
other.
It  follows  that  the  parity  transformation  leaves  the  type  (diagonal or
antidiagonal) of a fermionic state unchanged.
At  the  same   time,  it  is  known  that  one-dimensional  bound  states  are
nondegenerate \cite{LandauIII}.
By combining these facts, we can conclude that diagonal and  antidiagonal bound
states should possess certain parities.

We  first  consider  a   diagonal   fermionic  bound  state  $\psi_{\text{d}}$.
In Eq.~(\ref{III:14}), the argument  of  the  local Heun function tends to zero
as $x \rightarrow \infty$, and hence  the  local  Heun function tends to unity.
It follows that under the replacement $k \rightarrow i \kappa$, the transmitted
fermionic wave will have the correct bound state asymptotics $\propto\exp\left(
-\kappa x\right)$ as $x \rightarrow \infty$.
Under  the  replacement $k  \rightarrow  i \kappa$  in  Eq.~(\ref{III:16}), the
reflected  wave  also will have the  correct  asymptotics  $\propto \exp \left(
\kappa x\right)$ as $x\rightarrow -\infty$; however, in Eq.~(\ref{III:16}), the
incident wave  will  then  be $\propto \exp\left( -\kappa x \right)$  and  will
increase indefinitely as $x \rightarrow -\infty$.
To eliminate  the incorrect asymptotic  behavior of the incident fermionic wave
in Eq.~(\ref{III:16}), the coefficient $c_{1}\left(\varepsilon, k \right)$ must
vanish at  $\varepsilon = \varepsilon_{n},  k = i \kappa_{n} = i \left( M^{2} -
\varepsilon_{n}^{2} \right)^{1/2}$,  where $\varepsilon_{n}$ is the energy of a
diagonal bound state.
Note  that  the  coefficient $c_{1}\left(\varepsilon, k \right)$  is explicitly
determined by  Eq.~(\ref{III:18}) and  Eqs.~(\ref{III:20})--(\ref{III:23}).
Thus, the energy  levels  $\varepsilon_{n}$  of  the  diagonal fermionic  bound
states  are determined by the solutions of the transcendental equation
\begin{equation}
c_{1}\bigl(\varepsilon, i\left(M^{2}-\varepsilon^{2}\right)^{1/2}\bigr) = 0.
                                                                \label{III:46c}
\end{equation}
Similarly, we can conclude that the energy levels of the antidiagonal fermionic
bound states are determined  by the transcendental equation
\begin{equation}
d_{1}\bigl(\varepsilon, i\left(M^{2}-\varepsilon^{2}\right)^{1/2}\bigr) = 0,
                                                                \label{III:46d}
\end{equation}
where  the  coefficient  $d_{1}\left( \varepsilon, k \right)$   is   explicitly
determined by  Eq.~(\ref{III:24}) and  Eqs.~(\ref{III:26})--(\ref{III:28}).
Due  to  the  symmetry  of  the   Dirac   equation  under   charge  conjugation
(\ref{II:15a}), the energy levels of the  diagonal (antidiagonal) antifermionic
bound states are opposite  in  sign  to  those  of  the antidiagonal (diagonal)
fermionic bound states.

As in the case of fermion scattering,  the  wave  functions of the bound states
described above are the result of  matching   two   local analytic solutions of
Eq.~(\ref{III:8}) (or Eq.~(\ref{III:10})) at some intermediate point.
At the same time, there  are  bound states whose wave functions are analytic in
the entire complex plane, excluding the point at infinity.
To demonstrate this, we consider Eq.~(\ref{III:14}).
As  mentioned    above,   after   the   replacement $k  \rightarrow  i \kappa$,
Eq.~(\ref{III:14})  will  show  the  correct  asymptotic behavior $\propto \exp
\left(-\kappa x\right)$ as $x \rightarrow \infty$.
However, the  factor $\exp\left(-\kappa x \right)$  increases  indefinitely  as
$x \rightarrow -\infty$, whereas at large negative $x$, the correct asymptotics
of the bound state should be $\propto \exp\left(\kappa x \right)$.
Hence, to compensate for the unlimited growth of the factor $\exp \left(-\kappa
x\right)$ and to obtain the correct (i.e., $\propto \exp\left(\kappa x\right)$)
asymptotic  behavior  of  the   bound   state,   the   local  Heun  function in
Eq.~(\ref{III:14}) must  tend  to  zero as $\exp\left(2\kappa x\right)$, as $x
\rightarrow -\infty$.
The argument of the local Heun  function  entering  Eq.~(\ref{III:14}) tends to
unity as $x \rightarrow -\infty$.
It follows that the local Heun  function  should be analytic in the vicinity of
unity.

Recall that Heun's equation has four regular singular points at $z =0$, $z =a$,
$z = 1$, and  $z = \infty$.
The local Heun function $Hl\left(a, q; \alpha, \beta, \gamma, \delta; z\right)$
is analytic in the vicinity  of  the regular singular point $z = 0$, but in the
general case, it is not defined in the vicinity  of the regular singular points
$z = a$ and $z = 1$. 
However, if a local Heun function is analytic in  the  vicinity of $z = 0$  and
$z = 1$, it must also be analytic in  the  vicinity of $z = a = 1/2$, since the
regular singular point $z= a = 1/2$ lies between the other two regular singular
points $z = 0$ and $z = 1$.
The situation in which the local Heun function $Hl\left(a,q;\alpha,\beta,\gamma,
\delta; z\right)$ is  analytic  in  a  domain  containing  the  three  adjacent
singularities $z = 0$,  $z = a$,  and  $z = 1$ is rather specific.
In this case, $Hl\left(a, q; \alpha, \beta, \gamma, \delta; z \right)$  is also
a solution around  the fourth singularity $z  =  \infty$ and is reduced  to the
Heun polynomial \cite{Ronveaux, DLMF}.
A necessary condition for this is $\alpha= -n$, where $n$ is a positive integer.
Note that the local  Heun  functions  in  Eq.~(\ref{III:14}) have the parameter
$\alpha = -1$.

When $\alpha = -n$, where $n$  is  a  positive integer, the $(n + 1)\text{-th}$
coefficient in the  series  expansion  of $Hl\left(a, q; \alpha, \beta, \gamma,
\delta; z\right)$ is a polynomial in $q$ of order $n+1$.
If $q$ is a root of that polynomial, then the $(n + 1)$-th coefficient vanishes
and with it all the  following  ones, so the series is truncated and  the local
Heun function becomes the Heun polynomial. 
In our  case,   the   series   expansion   of   the   local   Heun   function in
Eq.~(\ref{III:14}) has the form
\begin{eqnarray}
&&Hl\left[ \frac{1}{2},-\frac{2\left( \kappa +i\varepsilon \right) }{m}
;-1,0,1+\frac{2\kappa }{m},1-\frac{2\kappa }{m};z\right]        \label{III:46a}
\\
&=&1-\frac{4i\left( \varepsilon -i\kappa \right) }{m+2\kappa }z-\frac{
4\varepsilon \left( \varepsilon -i\kappa \right) }{\left( m+\kappa \right)
\left( m+2\kappa \right)}z^{2}+O\left(z^{3}\right),   \nonumber
\end{eqnarray}
where the  parameter $\kappa  =  \left(M^{2}  -  \varepsilon^{2}\right)^{1/2}$.
We see that in Eq.~(\ref{III:46a}),  the  coefficient  of  $z^{2}$ vanishes iff
the energy $\varepsilon = 0$.
In this  case,  the  local  Heun  function  (\ref{III:46a})  is  reduced to the
first-order Heun polynomial $Hp_{1,1}\left[1/2,-2M/m;-1, 0, 1 + 2M/m, 1 - 2M/m;
z \right] = 1 - 4 z M/\left(m + 2 M \right)$.
The Heun polynomial vanishes at  the  regular  singular  point  $z = 1$ iff the
following condition holds
\begin{equation}
m = 2 M,                                                        \label{III:46b}
\end{equation}
which is equivalent to the condition
\begin{equation}
2 g = \lambda^{1/2}.                                            \label{III:46e}
\end{equation}

We conclude that in the external  field  of the sine-Gordon kink, the  diagonal
bound  state  with  zero  energy (diagonal  zero  mode)  exists  iff  condition
(\ref{III:46b}) holds.
Similarly, it can be shown  that  this  is  also true for the antidiagonal zero
mode.
In this case, the  analytic  wave  function  arises  from the reflected wave in
Eq.~(\ref{III:17}), and the local  Heun  function  entering  the reflected wave
also has the parameter $\alpha = -1$.

Let us denote the diagonal and  antidiagonal  zero  modes as $\psi_{0\text{d}}$
and $\psi_{0 \text{a}}$, respectively.
These zero  modes can then be written in the form
\begin{equation}
\psi _{0\text{d}}=\sqrt{\frac{2}{\pi i}}M^{1/2}
\begin{pmatrix}
-\dfrac{e^{M x}}{1 + i e^{2 M x}} & 0 \\
0 & \dfrac{e^{M x}}{i + e^{2 M x}}
\end{pmatrix},                                                   \label{III:47}
\end{equation}
\begin{equation}
\psi _{0\text{a}}=\sqrt{\frac{2}{\pi i}}M^{1/2}
\begin{pmatrix}
0 & \dfrac{e^{M x}}{i + e^{2 M x}} \\
\dfrac{e^{M x}}{1 + i e^{2 M x}} & 0
\end{pmatrix}.                                                   \label{III:48}
\end{equation}
The zero modes (\ref{III:47}) and (\ref{III:48}) are  normalized  to unity  and
become $\propto \exp\left( - M \left\vert x \right\vert \right)$ as $\left\vert
x\right\vert \rightarrow \infty$.

We  now  investigate  the  properties  of  the  zero  modes  (\ref{III:47}) and
(\ref{III:48})  under  the   charge  conjugation  (\ref{II:15a})   and   parity
transformation (\ref{II:15b}).
Choosing the phase factors $\eta_{C}$  and $\eta_{P}$  to  be  equal to $1$ and
$-1$, respectively, we find that
\begin{eqnarray}
\psi _{0\text{d}}^{C}\left( t,x\right)  &=&\psi _{0\text{a}}\left(
t,x\right),                                                      \label{III:49}
\\
\psi _{0\text{a}}^{C}\left( t,x\right)  &=&\psi _{0\text{d}}\left(
t,x\right),                                                      \label{III:50}
\\
\psi _{0\text{d}}^{P}\left( t,x\right)  &=&\psi _{0\text{d}}\left(
t,x\right),                                                      \label{III:51}
\\
\psi _{0\text{a}}^{P}\left( t,x\right)  &=&-\psi _{0\text{a}}\left(
t,x\right).                                                      \label{III:52}
\end{eqnarray}
We see that the diagonal and antidiagonal zero modes turn into each other under
the $C$-conjugation.
We also see that these zero modes  are  the  eigenstates of the parity operator
$P$, and that their eigenvalues (i.e., parities) are opposite.
Later, from our numerical  results,  we  shall  see  that  the antidiagonal and
diagonal  zero  modes  should  be  regarded  as  fermionic  and  antifermionic,
respectively.

Using the zero modes  $\psi_{0\text{d}}$  and  $\psi_{0\text{a}}$,  we can form
even and odd linear combinations:
\begin{eqnarray}
\psi_{0\text{e}}&=&2^{-1/2}\left(\psi_{0\text{d}}+
                   \psi_{0\text{a}}\right),                      \label{III:53}
\\
\psi_{0\text{o}}&=&i 2^{-1/2}\left(\psi_{0\text{d}}-
                   \psi_{0\text{a}}\right),                      \label{III:54}
\end{eqnarray}
which are the eigenstates of the $C$-conjugation operator
\begin{eqnarray}
\psi_{0\text{e}}^{C}\left( t,x\right) &=& \psi_{0\text{e}}
\left( t,x\right),                                               \label{III:55}
\\
\psi_{0\text{o}}^{C}\left( t,x\right) &=& \psi_{0\text{o}}
\left( t,x\right).                                               \label{III:56}
\end{eqnarray}
At the same time, the zero modes $\psi_{0\text{e}}$ and $\psi_{0\text{o}}$ turn
into each other up to a phase factor under the parity transformation
\begin{eqnarray}
\psi_{0\text{e}}^{P}\left(t,x\right)& = &-i \psi_{0\text{o}}\left(t,x\right),
                                                                 \label{III:57}
\\
\psi_{0\text{o}}^{P}\left(t,x\right)& = &i \psi_{0\text{e}}\left(t,x\right).
                                                                 \label{III:58}
\end{eqnarray}
Note also that the mean value of the  isospin  $x$-projection  vanishes for all
types of zero modes:
\begin{equation}
\left\langle I_{1}\right\rangle_{\psi _{0\text{d}}} =
\left\langle I_{1}\right\rangle _{\psi _{0\text{a}}} =
\left\langle I_{1}\right\rangle_{\psi _{0\text{e}}} =
\left\langle I_{1}\right\rangle_{\psi _{0\text{o}}} = 0.         \label{III:59}
\end{equation}

Eqs.~(\ref{III:55}) and  (\ref{III:56})  tell  us  that  $\psi_{0\text{e}}$ and
$\psi_{0\text{o}}$ are the Majorana spinors.
Using Eqs.~(\ref{III:47}), (\ref{III:48}),  (\ref{III:53}), and (\ref{III:54}),
it can be shown  that  $\psi_{0\text{e}}$  and  $\psi_{0\text{o}}$  satisfy the
relation
\begin{equation}
\bar{\psi}_{0\text{e}}\boldsymbol{\tau}_{\perp}\psi_{0\text{e}} =
\bar{\psi}_{0\text{o}}\boldsymbol{\tau}_{\perp}\psi_{0\text{o}} = 0.
                                                                \label{III:58a}
\end{equation}
It then follows  from  Eqs.~(\ref{II:8})  and (\ref{II:10}) that  the  Majorana
fermions corresponding  to  $\psi_{0\text{e}}$  and  $\psi_{0\text{o}}$ have no
effect on the field of the  kink, meaning that the external field approximation
becomes exact in this case.

Eqs.~(\ref{III:34a}) and (\ref{III:40a}) determine   the  phase  shifts  in the
scattering of the diagonal  and antidiagonal fermionic component, respectively.
These  phase  shifts  depend  on  the  magnitude  of  the  fermion's  momentum.
The difference in the phase shifts $\Delta_{\text{d,\,a}}=\delta_{\text{d,\,a}}
\left(0\right)  -  \delta_{\text{d,\,a}}\left(\infty\right)$ plays an important
role  in  the  theory  of  scattering  \cite{LandauIII, Goldberger, Taylor}; in
particular,  Levinson's   theorem  \cite{levinson_49}  establishes  a  relation
between this  difference  and the number of bound states for a given scattering
channel.
For a one-dimensional  case,  Levinson's  theorem  has  the form \cite{barton}:
\begin{equation}
\Delta = \pi \left(n_{b} - 1/2\right),                           \label{III:60}
\end{equation}
where $\Delta$ and $n_{b}$  are  the  difference  in  the  phase shifts and the
number of bound states in a given scattering channel, respectively.
In our case,  Levinson's theorem is written as
\begin{equation}
\Delta_{\text{d,\,a}} = \pi\left(n_{\text{d,\,a}} - 1/2 \right), \label{III:61}
\end{equation}
where $n_{\text{d}}$ $(n_{\text{a}})$ is the number of  diagonal (antidiagonal)
fermionic bound states.

Another well-known field  theory  kink  solution  is the kink of the $\phi^{4}$
model \cite{dhr_74, polyakov_jetpl_74, goldstone_jackiw_prd_75}.
Unlike the sine-Gordon kink, the $\phi^{4}$  kink  has  a  single Majorana zero
mode that exists for all values of the Yukawa coupling constant.
The presence of this zero mode leads  to  fragmentation of the fermionic charge
and polarization of the fermionic vacuum \cite{jackiw_rebbi_prd_76}.
We now consider the effect of the presence of two zero modes (\ref{III:47}) and
(\ref{III:48}) in the field of the sine-Gordon kink.
Recall that these zero modes exist iff condition (\ref{III:46b}) holds.
First, we note that  one can take either $\psi_{0\text{a}}$  and $\psi_{0\text{
d}}$ or $\psi_{0\text{e}}$ and $\psi_{0\text{o}}$ as the states of zero energy.
In the first case, the second quantized fermionic field is written as
\begin{eqnarray}
\Psi \left( t,x\right)  &=&b_{0}\psi _{0\text{a}}\left( x\right)
+d_{0}^{\dag }\psi _{0\text{d}}\left( x\right)                   \label{III:62}
\\
&&+\sum\limits_{r\geqslant 1}\left( b_{r}e^{-i\varepsilon _{r}t}\psi _{r
\text{d}}\left( x\right) +d_{r}^{\dag }e^{i\varepsilon _{r}t}\psi _{r\text{a}
}\left( x\right) \right),                                    \nonumber
\end{eqnarray}
where  it  is  understood   that   the   fermion-kink  system  is  placed in an
one-dimensional box of large but finite length, so  that  all  fermionic energy
levels are discrete.
In Eq.~(\ref{III:62}), the role of the diagonal  and antidiagonal zero modes is
reversed compared to the corresponding nonzero modes.
Our numerical results will show that all nonzero  diagonal (antidiagonal) modes
are fermionic (antifermionic) and that the situation is reversed for zero modes.
The annihilation and creation  operators  satisfy the anticommutation relations
\begin{equation}
\bigl\{ b_{r},b_{r^{\prime }}^{\dag }\bigr\} = \bigl\{ d_{r},d_{r^{\prime
}}^{\dag }\bigr\} = \delta _{rr^{\prime}},                       \label{III:63}
\end{equation}
where $r, r' = 0, 1, 2, \ldots\,.$
The other anticommutators involving  the  annihilation  and creation  operators
vanish.  

Next, to calculate the operator of  the  fermionic charge, we  use the normally
ordered fermionic current
\begin{equation}
j^{\mu} = 2^{-1}\left[\Psi ^{\dag },\gamma ^{0}\gamma ^{\mu }\Psi \right].
                                                                 \label{III:64}
\end{equation}
The corresponding expression for the fermionic charge operator is
\begin{eqnarray}
Q &=&2^{-1}\int \left( \Psi ^{\dag }\Psi -\Psi \Psi ^{\dag }\right) dx
                                                                 \label{III:65}
  \\
&=&b_{0}^{\dag }b_{0}-d_{0}^{\dag }d_{0}+\sum\limits_{r\geqslant 1}
\left(b_{r}^{\dag }b_{r}-d_{r}^{\dag }d_{r}\right).                \nonumber
\end{eqnarray}
It follows from  Eq.~(\ref{III:65})  that  the  contributions of the zero modes
$\psi_{0 \text{a}}$   and   $\psi_{0 \text{d}}$   to   the fermionic charge are
completely analogous to those of the nonzero modes.
Let us denote the state vectors in the subspace of  zero  modes  by $\left\vert
n_{\text{a}}, n_{\text{d}}\right\rangle_{(1)}$, where $n_{\text{a}\,(\text{d})}
=0, 1$ is the occupation number of the antidiagonal (diagonal) zero mode.
It then follows from Eqs.~(\ref{III:63}) and (\ref{III:65}) that
\begin{subequations}                                             \label{III:66}
\begin{eqnarray}
Q\left\vert 0,0\right\rangle _{(1)} &=&0,                       \label{III:66a}
 \\
Q\left\vert 1,0\right\rangle _{(1)} &=&\left\vert 1,0\right\rangle _{(1)},
                                                                \label{III:66b}
 \\
Q\left\vert 0,1\right\rangle _{(1)} &=&-\left\vert 0,1\right\rangle _{(1)},
                                                                \label{III:66c}
 \\
Q\left\vert 1,1\right\rangle _{(1)} &=&0.                       \label{III:66d}
\end{eqnarray}
\end{subequations}
We see that in the case of the zero modes $\psi_{0\text{a}}$ and $\psi_{0\text{
d}}$, there is no fermionic charge fragmentation, since all  eigenvalues of $Q$
are integers.
Furthermore, Eq.~(\ref{III:66a}) tells us  that  the  sine-Gordon kink does not
polarize the fermionic vacuum $\left\vert 0, 0 \right\rangle_{(1)}$ because its
fermionic charge vanishes.

Next, we consider the  case  where  zero  energy  fermions  are in the Majorana
states $\psi_{0\text{e}}$ and $\psi_{0\text{o}}$.
The second quantized fermionic field is then written as
\begin{eqnarray}
\Psi \left( t,x\right) &=&\alpha \psi_{0\text{e}}\left(x\right)+\beta
\psi_{0\text{o}}\left( x\right)                                  \label{III:67}
  \\
&&+\sum\limits_{r\geqslant 1}\left( b_{r}e^{-i\varepsilon _{r}t}\psi
_{r \text{d}}\left( x \right) + d_{r}^{\dag}e^{i\varepsilon_{r}t}
\psi_{r \text{a}}\left( x \right) \right),                      \nonumber
\end{eqnarray}
where the annihilation operators
\begin{subequations}                                             \label{III:68}
\begin{eqnarray}
\alpha  &=&2^{-1/2}\bigl( b_{0}+d_{0}^{\dag }\bigr),            \label{III:68a}
  \\
\beta  &=&i 2^{-1/2}\bigl( b_{0}-d_{0}^{\dag }\bigr)            \label{III:68b}
\end{eqnarray}
\end{subequations}
satisfy the anticommutation relations
\begin{equation}
\left\{ \alpha ,\alpha ^{\dag }\right\} = \left\{ \beta ,\beta ^{\dag
}\right\} = 1.                                                   \label{III:69}
\end{equation}
Note that Eq.~(\ref{III:67}) contains  only  the  annihilation operators of the
zero modes $\psi_{0\text{e}}$ and $\psi_{0\text{o}}$, and the creation operators
of the corresponding zero antimodes are absent.
This  is   because   the   wave   functions   of   the   Majorana   zero  modes
$\psi_{0\text{e}}$  and  $\psi_{0\text{o}}$  are  invariant  under  the  charge
conjugation, and hence the corresponding zero modes and antimodes are the same.

In terms of $\alpha$ and $\beta$, the fermionic charge $Q$ has the form
\begin{equation}
Q=\alpha ^{\dag }\alpha +\beta ^{\dag }\beta -1+\sum\limits_{r\geqslant
1}\left( b_{r}^{\dag }b_{r}-d_{r}^{\dag }d_{r}\right).           \label{III:70}
\end{equation}
We denote the state vectors  in the subspace of the Majorana zero modes $\psi_{
0\text{e}}$ and  $\psi_{0\text{o}}$  by  $\left\vert n_{\text{e}}, n_{\text{o}}
\right\rangle_{(2)}$.
The state  vectors  $\left\vert n_{\text{e}}, n_{\text{o}} \right\rangle_{(2)}$
are  linear  combinations  of  the  state vectors $\left\vert n_{\text{a}}, n_{
\text{d}}\right \rangle_{(1)}$:
\begin{subequations}                                             \label{III:71}
\begin{eqnarray}
\left\vert 0,0\right\rangle _{(2)} &=&\left\vert 0,1\right\rangle _{(1)},
                                                                \label{III:71a}
 \\
\left\vert 1,0\right\rangle _{(2)} &=&2^{-1/2}\bigl( \left\vert
0,0\right\rangle_{(1)}+\left\vert 1,1\right\rangle_{(1)}\bigr), \label{III:71b}
  \\
\left\vert 0,1\right\rangle _{(2)} &=&2^{-1/2}\bigl( \left\vert
0,0\right\rangle_{(1)}-\left\vert 1,1\right\rangle_{(1)}\bigr), \label{III:71c}
  \\
\left\vert 1,1\right\rangle _{(2)} &=&\left\vert 1,0\right\rangle _{(1)}.
                                                                \label{III:71d}
\end{eqnarray}
\end{subequations}
From Eqs.~(\ref{III:69}) and (\ref{III:70}), we obtain the relations:
\begin{subequations}                                             \label{III:72}
\begin{eqnarray}
Q\left\vert 0,0\right\rangle_{(2)} &=&-\left\vert 0,0\right\rangle_{(2)},
                                                                \label{III:72a}
 \\
Q\left\vert 1,0\right\rangle _{(2)} &=&0,                       \label{III:72b}
 \\
Q\left\vert 0,1\right\rangle _{(2)} &=&0,                       \label{III:72c}
 \\
Q\left\vert 1,1\right\rangle_{(2)} &=&\left\vert 1,1\right\rangle_{(2)}.
                                                                \label{III:72d}
\end{eqnarray}
\end{subequations}
We see that as in the previous case, there is no fermionic charge fragmentation.
At the same time, it follows from Eq.~(\ref{III:72a}) that the fermionic charge
of the vacuum state is  equal  to  minus  one,  and  hence the sine-Gordon kink
polarizes the fermionic vacuum $\left\vert 0, 0 \right\rangle_{(2)}$.

\section{\label{sec:V} Numerical results}

It follows from Eqs.~(\ref{III:1}) and (\ref{III:2})  that  the Dirac equations
for the diagonal and  antidiagonal  components  of  the fermionic wave function
contain  the  meson  mass  $m$  and  fermion  mass  $M = m g \lambda^{-1/2}$ as
parameters.
Passing to the dimensionless variable $\tilde{x}= m x$, it is easy to show that
the solutions to the Dirac equation depend only on the dimensionless  variables
$\tilde{x}= m x$, $\tilde{M} = M/m$, $\tilde{\varepsilon} = \varepsilon/m$, and
$\tilde{k} = k/m$, in accordance with Eqs.~(\ref{III:14}) -- (\ref{III:17}).
Hence, the meson mass $m$ can be taken equal to unity in numerical calculations,
while the calculated values can be presented as functions  of the dimensionless
fermion momentum $\tilde{k} = k/m$ or mass $\tilde{M} = M/m$.
In terms of these dimensionless  variables,  condition (\ref{III:III}) ensuring
the applicability of the external field approximation takes the form
\begin{equation}
\tilde{g} \ll 8 \tilde{\lambda}^{-1/2},                             \label{V:I}
\end{equation}
where $\tilde{g} = g/m$ and $\tilde{\lambda} = \lambda/m^{2}$.
From Eq.~(\ref{V:I}), it follows that the dimensionless fermion mass $\tilde{M}
=\tilde{g}\tilde{\lambda}^{-1/2}$ must be much less than the value of $8\tilde{
\lambda}^{-1}= 8 \times 10^{2}$, since we used $\tilde{\lambda}=10^{-2}$ in the
numerical calculations.

\begin{figure}[tbp]
\includegraphics[width=0.5\textwidth]{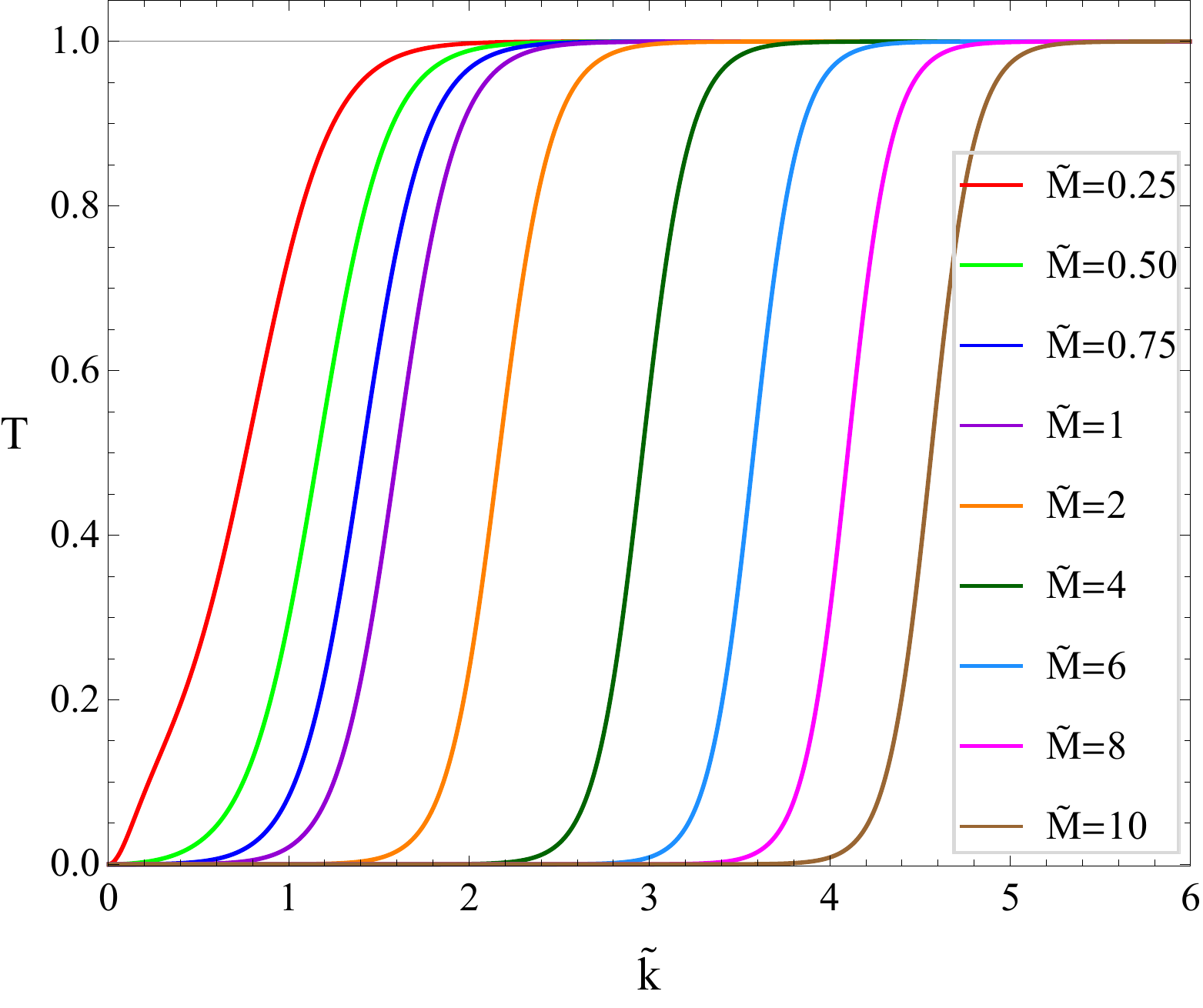}
\caption{\label{fig1}   Dependences of the transmission coefficient $T$ for the
antidiagonal component of  the fermionic  wave function on the fermion momentum
$\tilde{k}$ for different values of the fermion mass $\tilde{M}$.}
\end{figure}

\begin{figure}[tbp]
\includegraphics[width=0.5\textwidth]{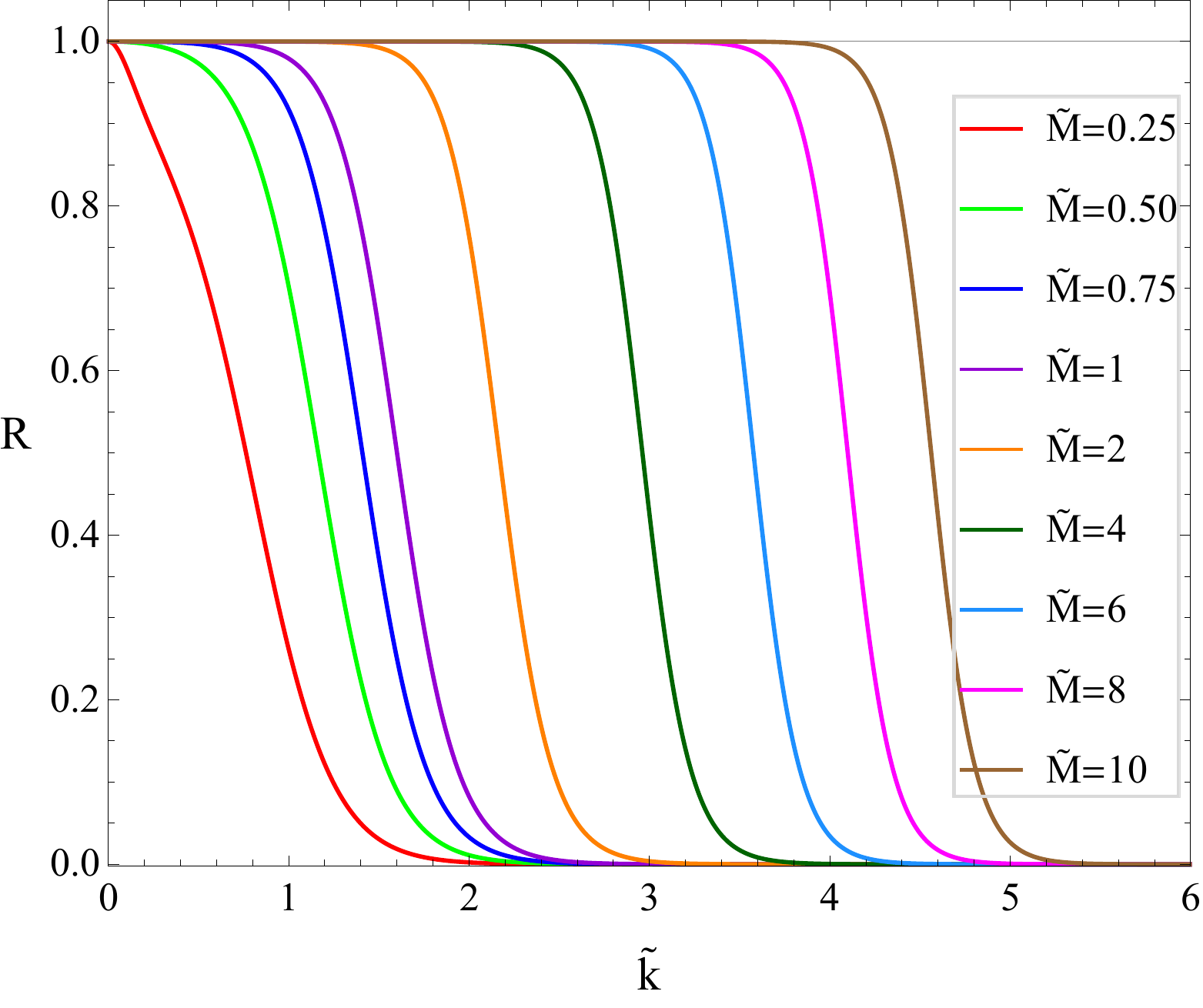}
\caption{\label{fig2}    Dependences  of the reflection coefficient $R$ for the
antidiagonal component of  the fermionic  wave function on the fermion momentum
$\tilde{k}$ for different values of the fermion mass $\tilde{M}$.}
\end{figure}

Figure~\ref{fig1} presents  the  dependences  of the transition coefficient $T$
for  the  antidiagonal  component  of   the  fermionic  wave  function  on  the
dimensionless fermion momentum $\tilde{k}$.
These dependences   were  obtained   using   the   analytical   expressions  in
Sec.~\ref{sec:III}, and are  shown  for  different  values of the dimensionless
fermion mass $\tilde{M}$.
We see that the curves $T(\tilde{k})$  have  a  characteristic tanh-like shape.
As the fermion mass $\tilde{M}$  increases,  the  curves monotonically shift to
the region of larger fermion momenta.
Figure~\ref{fig2} shows the curves $R(\tilde{k})$ that correspond to the curves
$T(\tilde{k})$ in Fig.~\ref{fig1}.
It was found that for the same values of $\tilde{M}$, the curves $T(\tilde{k})$
and $R(\tilde{k})$ satisfy the unitarity condition
\begin{equation}
T(\tilde{k}) + R(\tilde{k}) = 1.                                   \label{IV:1}
\end{equation}

\begin{figure}[tbp]
\includegraphics[width=0.5\textwidth]{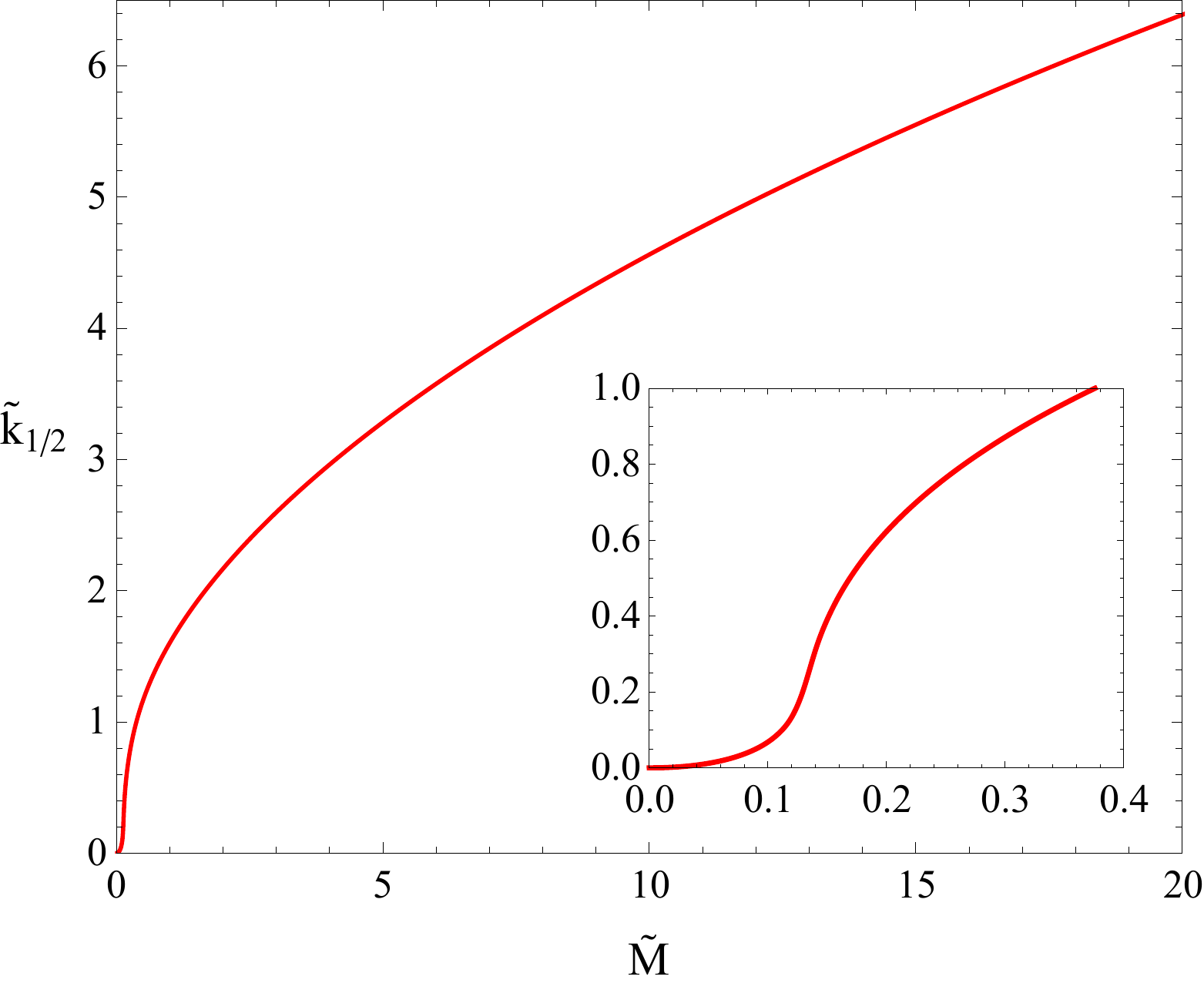}
\caption{\label{fig3}   Dependence of  the  parameter  $\tilde{k}_{1/2}$ on the
fermion mass $\tilde{M}$ for the antidiagonal case.}
\end{figure}

Let us define the fermion momentim $\tilde{k}_{1/2}$  by  means of the relation
$T(\tilde{k}_{1/2}) = R(\tilde{k}_{1/2}) = 1/2$.
For a given value of $\tilde{M}$, the value of $\tilde{k}_{1/2}$ determines the
position of the  midpoints  of  the curves $T(\tilde{k})$ and $R(\tilde{k})$ in
Figs.~\ref{fig1} and \ref{fig2}, respectively.
Hence, a  fermion  with  momentum  $\tilde{k}_{1/2}$  passes  through  the kink
barrier with probability $1/2$.
Figure~\ref{fig3}    presents   the   dependence   $\tilde{k}_{1/2}(\tilde{M})$
corresponding to the curves shown in Figs.~\ref{fig1} and \ref{fig2}.
It was found  numerically  that  $\tilde{k}_{1/2} \approx 4.2\tilde{M}^{2}$ for
$\tilde{M} \lesssim 0.1$, and $\tilde{k}_{1/2}\approx 1.43 \tilde{M}^{1/2}$ for
$\tilde{M} \gtrsim 1$.
Hence, the  fermion  velocity $v_{1/2} \approx \tilde{k}_{1/2}/\tilde{M} < 0.1$
if $\tilde{M} < 0.024$.
We see that in the antidiagonal case,  the  fermions that pass through the kink
barrier with  probability  $1/2$  are  nonrelativistic  when  the  fermion mass
$\tilde{M}$ is small.
The fermions also become nonrelativistic  for large fermion masses $\tilde{M} >
205$, as in this case,  the  fermion  velocity $v_{1/2}\approx 1.43 \tilde{M}^{
-1/2} < 0.1$.
In the  intermediate  mass  region,  the  fermions are moderately relativistic.

\begin{figure}[tbp]
\includegraphics[width=0.5\textwidth]{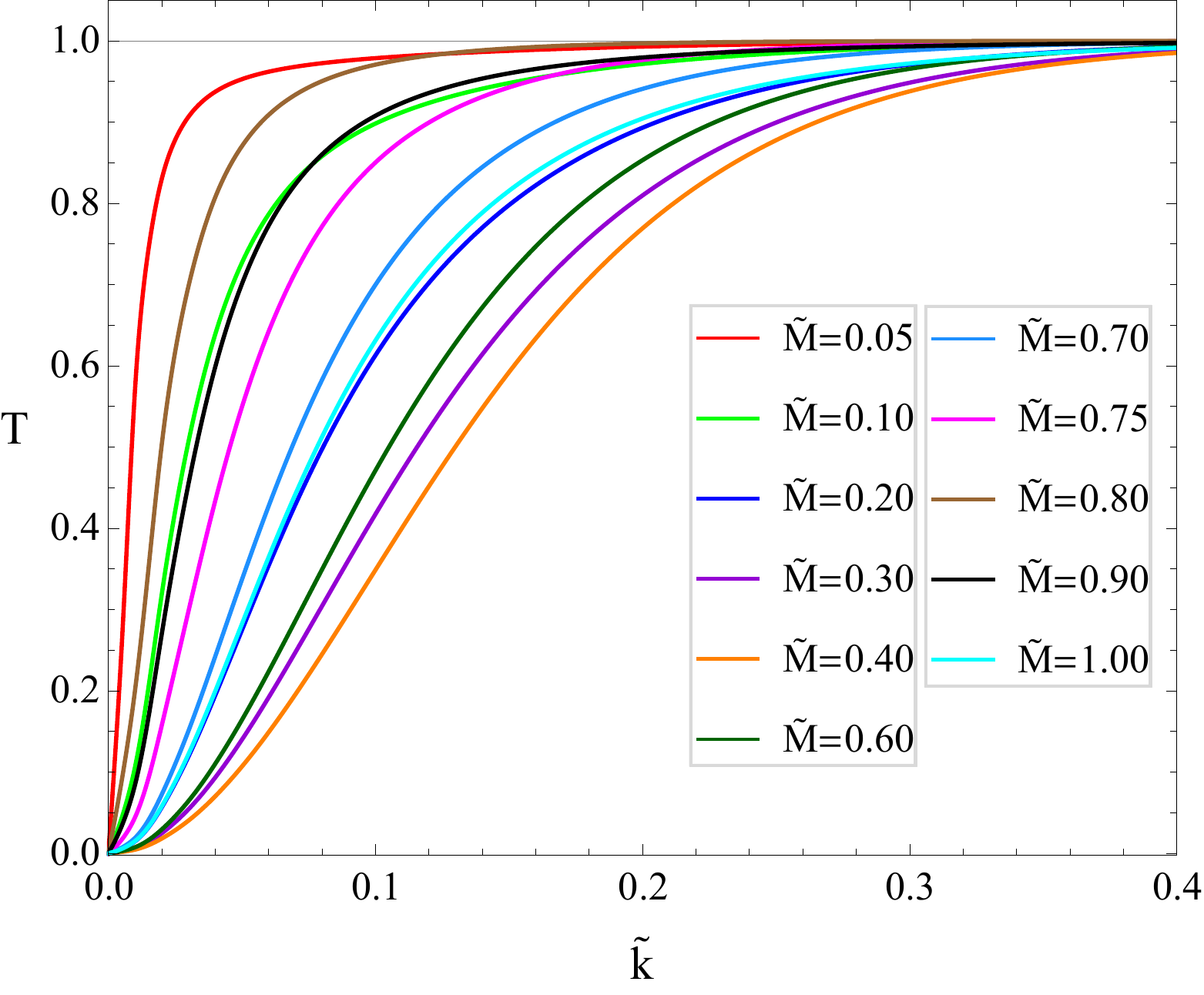}
\caption{\label{fig4}   Dependences of the transmission coefficient $T$ for the
diagonal component  of  the  fermionic  wave  function  on the fermion momentum
$\tilde{k}$ for different values of the fermion mass $\tilde{M}$.}
\end{figure}

\begin{figure}[tbp]
\includegraphics[width=0.5\textwidth]{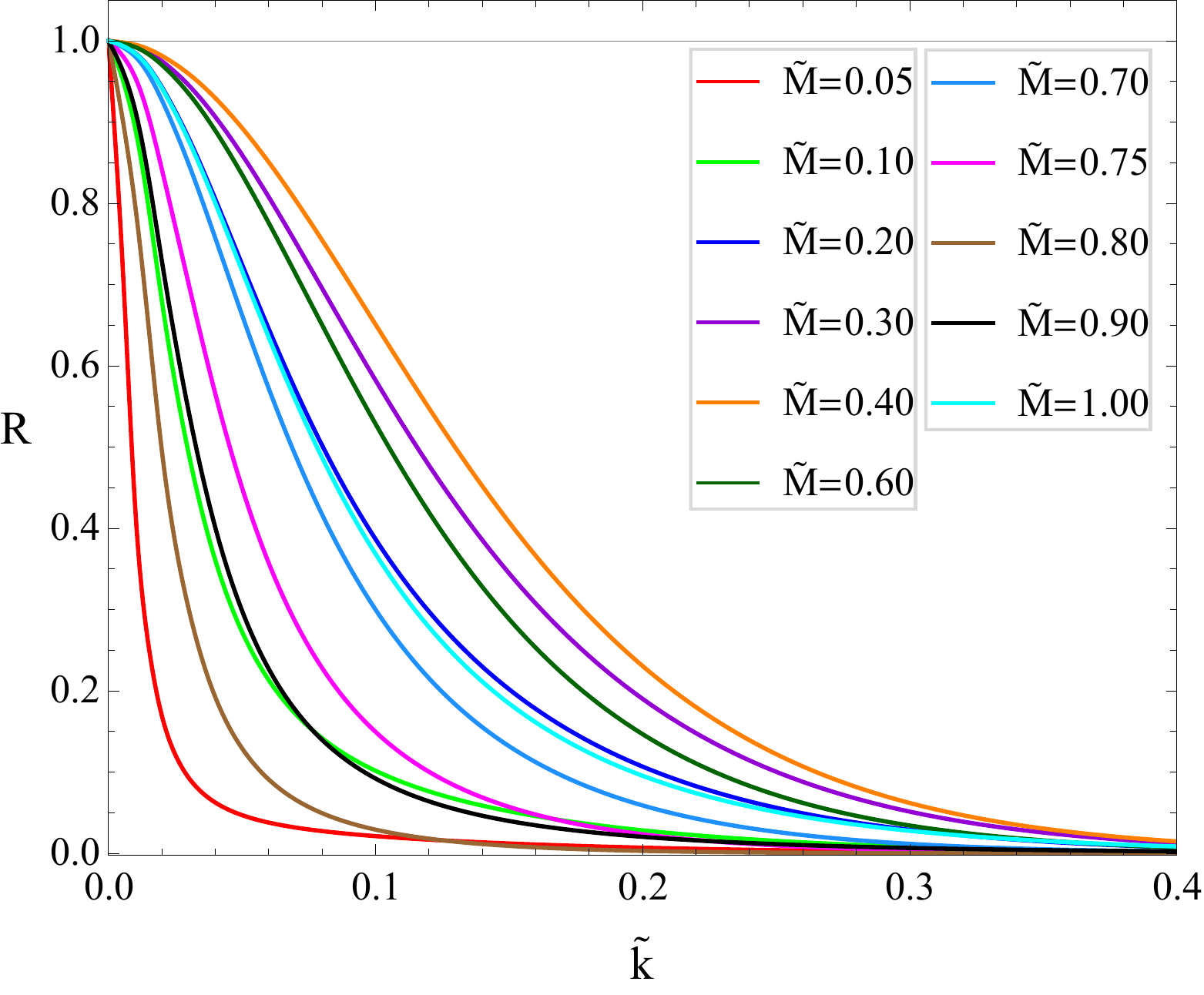}
\caption{\label{fig5}    Dependences  of the reflection coefficient $R$ for the
diagonal  component  of  the fermionic  wave  function  on the fermion momentum
$\tilde{k}$ for different values of the fermion mass $\tilde{M}$.}
\end{figure}

Next, we  turn  to  the  diagonal  component  of  the  fermionic wave function.
In Figs.~\ref{fig4} and \ref{fig5} we  can  see  the  curves $T(\tilde{k})$ and
$R(\tilde{k})$, respectively.
These curves carry information about  the interaction of the diagonal fermionic
component with the sine-Gordon kink.
As in the previous case,  the  curves  in Figs.~\ref{fig4} and \ref{fig5} which
correspond to the same  $\tilde{M}$  satisfy  unitarity condition (\ref{IV:1}).
However, except for this aspect, the behavior of the curves in Figs.~\ref{fig4}
and  \ref{fig5}  differs  substantially  from  those  in  Figs.~\ref{fig1}  and
\ref{fig2}, respectively.
In  the  antidiagonal   case  (Figs.~\ref{fig1}  and  \ref{fig2}),  the  curves
$T(\tilde{k})$ and $R(\tilde{k})$ monotonically shift to larger fermion momenta
with an increase in the fermion mass $\tilde{M}$.
In contrast, in the  diagonal case (Figs.~\ref{fig4} and \ref{fig5}), the shift
of the curves $T(\tilde{k})$ and $R(\tilde{k})$  has  an  oscillatory character
as $\tilde{M}$ increases.

\begin{figure}[tbp]
\includegraphics[width=0.5\textwidth]{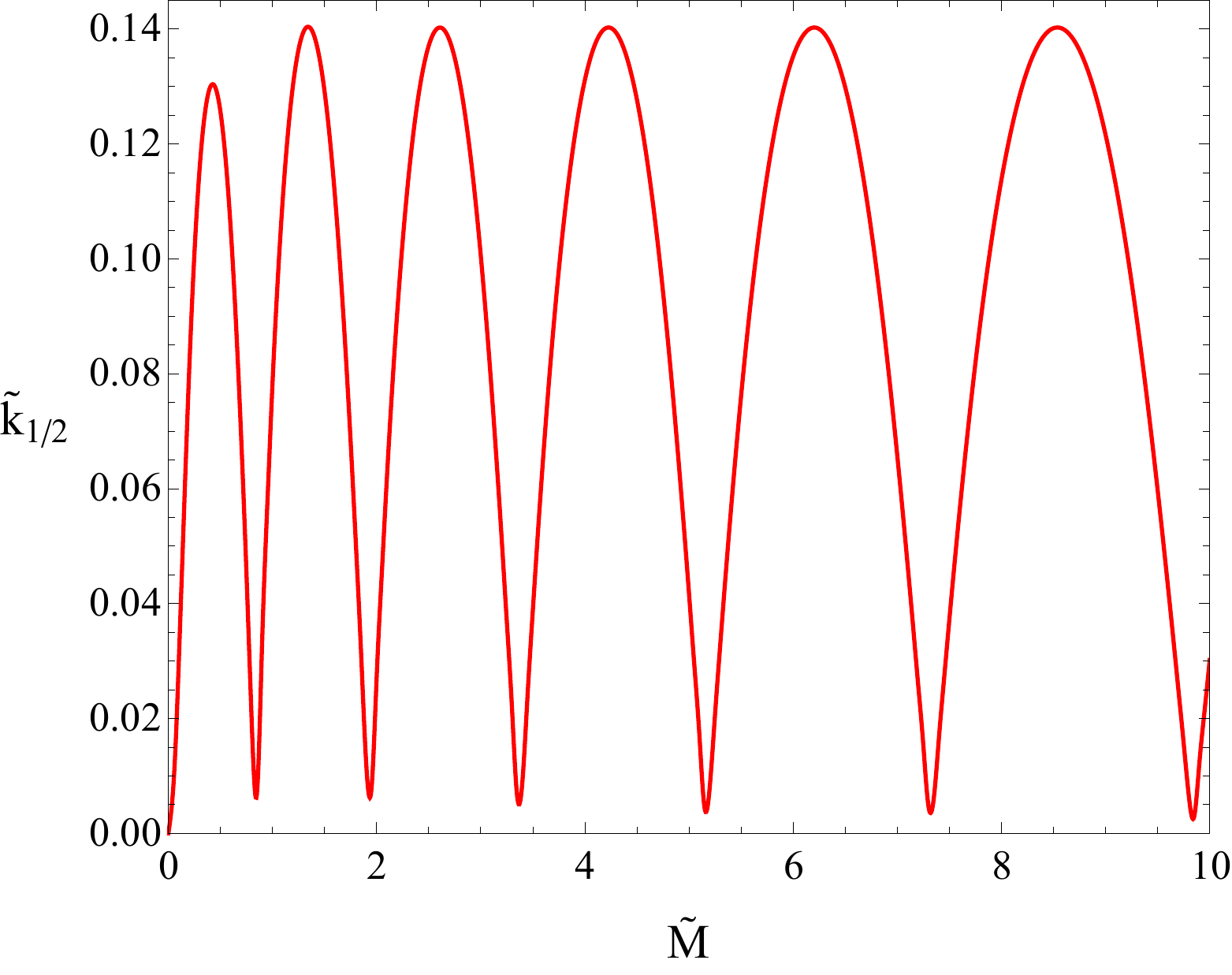}
\caption{\label{fig6}    Dependence of  the  parameter $\tilde{k}_{1/2}$ on the
fermion mass $\tilde{M}$ for the diagonal case.}
\end{figure}

Figure~\ref{fig6}    shows    the    dependence    $\tilde{k}_{1/2}(\tilde{M})$
corresponding to the curves in Figs.~\ref{fig4} and \ref{fig5}.
We see  that  the  dependences  $\tilde{k}_{1/2}( \tilde{M} )$  are  completely
different in the antidiagonal and diagonal cases.
The  curve   $\tilde{k}_{1/2}(\tilde{M})$   monotonically   increases   in  the
antidiagonal case in  Fig.~\ref{fig3}, whereas it has an  oscillatory character
in the diagonal case in  Fig.~\ref{fig6}.
The amplitude of the oscillations is  approximately $0.14$, and the oscillation
period increases with the fermion mass $\tilde{M}$.
In particular, it was found that for $n \gtrsim 3$, the  position $\tilde{M}_{n
}$ of  the  $n$-th  minimum of the curve $\tilde{k}_{1/2}(\tilde{M})$ satisfies
the asymptotic quadratic relation
\begin{equation}
\tilde{M}_{n} \sim a n^{2},                                        \label{IV:2}
\end{equation}
where the coefficient $a \approx 0.18$.
It follows that  the  oscillation  period (i.e., the  distance between adjacent
minima) grows linearly with $i$:
\begin{equation}
\Delta \tilde{M}_{n} = \tilde{M}_{n} - \tilde{M}_{n-1}\sim 2 a n.  \label{IV:3}
\end{equation}

From Eqs.~(\ref{III:34a}) and (\ref{III:40a}), we can find numerical values for
the phase shifts for the antidiagonal and diagonal fermionic scattering states.
It was found that in the antidiagonal case, the phase shift $\Delta_{\text{a}}$
does not  depend on $\tilde{M}$ and is equal to $\pi/2$.
In  contrast,  in  the  diagonal  case,  the  phase  shift  $\Delta_{\text{d}}$
increases step-wise by $\pi$ when the  fermion mass $\tilde{M}$ passes the next
minimum in the  curve  $\tilde{k}_{1/2}(\tilde{M})$  shown  in Fig.~\ref{fig6}.
Thus, the dependences of the  phase  shifts on the fermion mass $\tilde{M}$ are
described by the expressions:
\begin{eqnarray}
\Delta_{\text{a}}\bigl( \tilde{M}\bigr)  &=&\frac{\pi }{2},       \label{IV:4a}
\\
\Delta_{\text{d}}\bigl( \tilde{M}\bigr)  &=&\pi \biggl[ \frac{1}{2}
+ \sum_{n = 1}^{\infty}\theta \bigl(\tilde{M} - \tilde{M}_{n}\bigr) \biggr],
                                                                  \label{IV:4b}
\end{eqnarray}
where $\theta$ is the Heaviside function and $\tilde{M}_{1}\approx 0.85$ is the
minimum of the function $\tilde{k}_{1/2}(\tilde{M})$ in Fig.~\ref{fig6} that is
nearest to $\tilde{M} = 0$.
Levinson's  theorem (\ref{III:61}) and  Eq.~(\ref{IV:4a}) tell  us that for all
values of the fermion mass $\tilde{M}$, there is only one antidiagonal fermionic
bound state in the external field of the sine-Gordon kink.
On  the  other  hand,   according   to   Levinson's  theorem (\ref{III:61}) and
Eq.~(\ref{IV:4b}), the  number  of  diagonal  fermionic  bound states increases
step-wise  with an increase in $\tilde{M}$,  and  there  is always at least one
diagonal fermionic  bound  state in the external field of the sine-Gordon kink.

In order to verify the validity of these statements, we need to investigate the
spectrum of bound states of the Dirac Hamiltonian.
It is sufficient to study the  spectrum  of  the  antidiagonal component of the
Dirac  Hamiltonian,  since  for  a  given  value  of  $\tilde{M}$,  the  energy
eigenvalues of the diagonal bound  states  are opposite in sign to those of the
antidiagonal bound states.
To calculate the  energy  eigenvalues  of  the  antidiagonal  bound  states, we
used the numerical methods provided  by the {\sc{Mathematica}} software package
\cite{Mathematica}.
Figure~\ref{fig7} presents  the  dependence  of  the  ratio $\tilde{\varepsilon
}/\tilde{M}$  for the antidiagonal bound states of the Dirac Hamiltonian on the
fermion mass $\tilde{M}$.
The corresponding curves for the diagonal  bound states are obtained from those
shown in Fig.~\ref{fig7}  by  reflection  with  respect  to the horizontal zero
axis.

Let us examine the main features of the curves shown in Fig.~\ref{fig7}.
Firstly, we  see  that  there  is  only  one antidiagonal fermionic bound state
in the entire interval of fermion masses $\tilde{M}$.
When $\tilde{M} \equiv M/m = 1/2$, the energy  of this bound state is zero, and
thus the sine-Gordon kink  possesses  a  fermionic  zero  mode  as described in
Sec.~\ref{sec:IV}.
Conversely,  we  see  that  as  the  fermion  mass  $\tilde{M}$  increases, new
antidiagonal antifermionic  bound  states  arise  in  the external field of the
sine-Gordon kink, and one antidiagonal antifermionic bound state exists for all
values of $\tilde{M}$.
The situation is reversed for the diagonal bound states.
In this case, there is only  one  diagonal  antifermionic  bound state, whereas
new diagonal fermionic bound states arise as $\tilde{M}$ increases.
When $\tilde{M} \equiv M/m = 1/2$, the  energy  of  the  diagonal antifermionic
bound state vanishes, and it turns into the antifermionic zero mode.
The coordinates  $\tilde{M}_{n}$  of  the  points  at  which  new  antidiagonal
antifermionic  (diagonal  fermionic)  bound  states  arise  from  the continuum
coincide with the coordinates of the corresponding minima of the curve $\tilde{
k}_{1/2}(\tilde{M})$ in Fig.~\ref{fig6}.
It was found that in Fig.~\ref{fig7}, the curves $\left\vert \tilde{\varepsilon
}_{n}\right\vert/\tilde{M}$ tend quadratically to unity as $\tilde{M}$ tends to
$\tilde{M}_{n}$ from the right:
\begin{equation}
\frac{\left\vert \tilde{\varepsilon}_{n}\right\vert }{\tilde{M}} \approx
1 - \beta_{n} \bigl(\tilde{M} - \tilde{M}_{n}\bigr)^{2},           \label{IV:5}
\end{equation}
where $\beta_{n}$  are  some positive  constants  and  the  index $n = 0, 1, 2,
\ldots$ enumerates the curves in order of increasing $\tilde{M}_{n}$.
A qualitative explanation  of  this  behavior of the curves $\left\vert \tilde{
\varepsilon}_{n}\right\vert/\tilde{M}$ is given in the Appendix B.

It follows  from  the  above  that  Eqs.~(\ref{IV:4a}), (\ref{IV:4b}),  and the
the behavior of the curves in  Figs.~\ref{fig3}, \ref{fig6}, and \ref{fig7} are
in  agreement  with  Levinson's theorem.
Indeed, it follows from Eq.~(\ref{IV:4b}) that in the scattering of the diagonal
state fermions, the  phase  shift  $\Delta_{\text{d}}$  increases  step-wise by
$\pi$  when  $\tilde{M}$  passes  the  point  $\tilde{M}_{n}$  at  which  a new
diagonal fermionic bound state arises from the continuum.
At the same time, in Fig.~\ref{fig6},  the  curve  $\tilde{k}_{1/2}(\tilde{M})$
reaches a local near-zero  minimum  at $\tilde{M} = \tilde{M}_{n}$.
This corresponds  to  the  fact  that  diagonal  state fermions with $\tilde{M}
\approx \tilde{M}_{n}$ and small momenta $\tilde{k}$ almost completely transmit
through the kink barrier.
This  resonance  behavior  is  due  to  the  presence  of  a  virtual  level at
$\tilde{k}  \approx  0$  when  $\tilde{M}  \approx  \tilde{M}_{n}$,  and  is in
accordance   with    the    general    principles     of    scattering   theory
\cite{LandauIII, Goldberger, Taylor}.

In contrast, it follows from Eq.~(\ref{IV:4a}) that  in  the  scattering of the
antidiagonal state  fermions,  the  phase  shift  $\Delta_{\text{a}}$  does not
depend  on $\tilde{M}$ and is equal to $\pi/2$.
From Fig.~\ref{fig3}, we  see  that  the  corresponding  curve $\tilde{k}_{1/2}
(\tilde{M})$ increases monotonically with an increase in $\tilde{M}$ and has no
local minima at nonzero $\tilde{M}$.
Hence, the scattering of  the  antidiagonal  state  fermions on the sine-Gordon
kink does not have a resonance character.
This corresponds to the fact that  in  the  external  field  of the sine-Gordon
kink, there are no antidiagonal fermionic states arising at nonzero $\tilde{M}$
and there is only one such state arising at $\tilde{M} = 0$.
In addition, we note that the roles of the diagonal and antidiagonal states are
reversed when passing from fermions to antifermions.
The  diagonal  (antidiagonal)  antifermionic   states   are  scattered  on  the
sine-Gordon kink  in  the  same  way  as  the  antidagonal (diagonal) fermionic
states.

\section{\label{sec:VI} Conclusion}

In the  present  paper,   fermion  scattering  in  the  background field of the
sine-Gordon kink  has  been investigated  both  analytically  and  numerically.
To achieve  symmetry   of    the    fermion-kink    interaction  under discrete
transformation (\ref{II:3}), we  treat  the  sine-Gordon  model  as a nonlinear
$\sigma$-model with a circular target  space,  which interacts with a fermionic
isodoublet through the Yukawa interaction.
It was found that with respect to its  spin  and isospin indices, the fermionic
isodoublet can be divided into diagonal and  antidiagonal  parts  that interact
with the sine-Gordon kink independently of each other.

Studying the fermion-kink scattering, we  have found analytical expressions for
the wave functions of the diagonal and  antidiagonal fermionic states, and have
shown that these wave functions can be expressed in terms of the Heun functions.
Using the expressions obtained in this way and  the matching conditions for the
fermionic wave functions, we have derived general expressions for the fermionic
transmission and reflection coefficients.
It was found that the  scattering  of  the  diagonal  fermionic  states differs
significantly from that of the antidiagonal states.
In particular,  for  the  diagonal  fermionic  states,  the  dependence  of the
transmission and reflection coefficients on the fermion mass has an oscillatory
resonance character.
In contrast, this dependence has  a  monotonic  nonresonance  character for the
antidiagonal fermionic states.
For antifermions the situation is reversed: the scattering  of the antidiagonal
(diagonal) antifermionic  states  has   a  resonance  (nonresonance) character.

The fermion-kink system has a rather interesting structure of the bound states.
Their energy levels are determined by  zeros of the coefficient of the incident
wave in the asymptotics of the fermionic scattering states.
The number of bound  states  increases  as  the  Yukawa  coupling  $g$ constant
increases, which is equivalent to an increase in the fermion mass $M$.
The growth in the number of bound states  is  asymptotically $\propto M^{1/2} =
m^{1/2} g^{1/2}\lambda^{-1/4}$.
At the same time, for any nonzero $M$, there  are  at  least four bound states,
of which two are fermionic and two are antifermionic.
The diagonal fermionic (antifermionic)  bound  states are related by the charge
conjugation to the antidiagonal antifermionic (fermionic) bound states, meaning
that the energy levels of the  fermion-kink  system  can  be divided into pairs
of levels with opposite energies.

In addition,  the  fermion-kink  system  will  possesses  two  zero  modes when
condition (\ref{III:46b}) is satisfied.
Of these, the  antidiagonal zero mode is fermionic, while the  diagonal  one is
antifermionic, and these modes are related by the charge conjugation.
The two linear combinations of the  fermionic  and antifermionic zero modes are
the eigenstates of the charge conjugation  operator, and hence are the Majorana
zero modes.
The Majorana zero modes  have  no  effect  on  the  field  configuration of the
sine-Gordon kink.

It should be noted that as the fermion  mass  $M$  increases, the number of the
antidiagonal fermionic (diagonal antifermionic) states  remains  equal  to one,
while the number of  the antidiagonal antifermionic (diagonal fermionic) states
increases asymptotically $\propto M^{1/2}$.
This difference in the properties  of  the  bound  states appears to be closely
related through Levinson's  theorem  to  the  above-mentioned difference in the
scattering of the  diagonal  and  antidiagonal  states on the sine-Gordon kink.

In addition to the sine-Gordon kink, there  is  the well-known kink solution of
the  $(1  +  1)$-dimensional $\phi^{4}$  model \cite{dhr_74, polyakov_jetpl_74,
 goldstone_jackiw_prd_75}.
The $\phi^{4}$ kink possesses  a  single  Majorana  zero mode, which exists for
all nonzero values of the Yukawa coupling constant.
This is the main  difference  from  the  sine-Gordon  kink,  which has two zero
modes, but only if condition (\ref{III:46b}) is satisfied.
As in the previous case, the Majorana zero  mode  has  no  effect  on the field
configuration  of the $\phi^{4}$ kink.
The presence of the single Majorana zero  mode  leads  to  fragmentation of the
fermionic charge and polarization  of  the  fermionic  vacuum  in  the external
field of the $\phi^{4}$ kink, for  any  nonzero  value  of  the Yukawa coupling
constant.
In contrast, the presence of the two zero modes $\psi_{0\text{a}}$ and $\psi_{0
\text{d}}$ in the external field of the  sine-Gordon  kink makes it possible to
form the two Majorana  zero  modes $\psi_{0\text{e}}$ and $\psi_{0\text{o}}$ in
the exceptional case (\ref{III:46b}).
The presence of  two  zero  modes  makes  fragmentation of the fermionic charge
impossible in  the  external  field  of  the  sine-Gordon  kink.
At the same time, the  fermionic  charge  of  the vacuum state $\left\vert 0, 0
\right\rangle_{(2)}$ is equal to  minus  one,  which  indicates polarization of
the fermionic vacuum in the subspace of the Majorana zero modes.

As in the case of the sine-Gordon kink, the number of fermionic bound states of
the $\phi^{4}$ kink increases with an increase in the fermionic mass $M$.
However,  the  number  of  bound  fermionic  states  of  the $\phi^{4}$ kink is
asymptotically $\propto M$, while that of the sine-Gordon kink is asymptotically
$\propto M^{1/2}$.
Hence, we can say that the $\phi^{4}$ kink holds fermions more efficiently than
the sine-Gordon kink.

%
%

\appendix

\section{Plane-wave solutions to the Dirac equation}

When $\left\vert x \right\vert \gg m^{-1}$,  the   Dirac  equation (\ref{II:9})
describing fermions in external  kink  field  (\ref{II:12}) turns into the free
Dirac equation
\begin{equation}
i\gamma ^{\mu }\otimes \mathbb{I}\partial_{\mu }\psi - M \mathbb{I} \otimes
\tau_{1}\psi = 0,                                                   \label{A:1}
\end{equation}
where the fermion mass  $M = m g \lambda^{-1/2}$, and  we  explicitly write out
the matrices acting on the spinor and isospinor indices.
The free Dirac Hamiltonian corresponding to Eq.~(\ref{A:1}) is
\begin{equation}
H_{0}=\alpha \mathbb{\otimes I}\left( - i \partial_{x}\right) +
M \beta \mathbf{\otimes}\tau _{1},                                  \label{A:2}
\end{equation}
where $\alpha = \gamma^{0}\gamma^{1}$ and $\beta = \gamma^{0}$.
The Hamiltonian (\ref{A:2}) commutes with the operators $p_{x}=-i \partial_{x}$
and $I_{1} = 2^{-1}\tau_{1}$  corresponding   to   the   momentum  and  isospin
$x$-projection of a free fermion, respectively.
It follows  that  the  states  of  free  fermions  can  be characterized by the
momentum $p_{x}$ and isospin $x$-projection $I_{1}$.
Since in the  one-dimensional  case  fermions  have  no  spin, there exist four
fermionic states at a fixed fermion momentum $p_{x}$.
These states correspond  to  the  combinations  of  the isospin $x$-projections
$I_{1}=\pm 1/2$ and the fermion energies $p_{0}=\pm \varepsilon= \pm (p_{x}^{2}
+ M^{2})^{1/2}$ with both signs.

Let us denote the wave  function  of  a  fermion  with momentum $p_{x}$, energy
$p_{0}=\epsilon\left(p_{x}^{2} + M^{2}\right)^{1/2}$, where $\epsilon = \pm 1$,
and isospin $x$-projection  $I_{x}$  by $\psi_{p_{x}, \epsilon, I_{1}}$.
Then, the wave functions of free fermions have the form
\begin{widetext}
\begin{subequations}                                                \label{A:3}
\begin{eqnarray}
\psi _{p_{x},1,1/2} &=&\frac{1}{2\sqrt{\varepsilon L}}
\begin{pmatrix}
\sqrt{\varepsilon +p_{x}} & \sqrt{\varepsilon +p_{x}} \\
\sqrt{\varepsilon -p_{x}} & \sqrt{\varepsilon -p_{x}}
\end{pmatrix}
\exp \left[ -i\left( \varepsilon t-p_{x}x\right) \right],          \label{A:3a}
\\
\psi _{p_{x},-1,1/2} &=&\frac{1}{2\sqrt{\varepsilon L}}
\begin{pmatrix}
-\sqrt{\varepsilon -p_{x}} & -\sqrt{\varepsilon -p_{x}} \\
\sqrt{\varepsilon +p_{x}} & \sqrt{\varepsilon +p_{x}}
\end{pmatrix}
\exp \left[ i\left( \varepsilon t+p_{x}x\right) \right],           \label{A:3b}
\\
\psi _{p_{x},1,-1/2} &=&\frac{1}{2\sqrt{\varepsilon L}}
\begin{pmatrix}
-\sqrt{\varepsilon +p_{x}} & \sqrt{\varepsilon +p_{x}} \\
\sqrt{\varepsilon -p_{x}} & -\sqrt{\varepsilon -p_{x}}
\end{pmatrix}
\exp \left[ -i\left( \varepsilon t-p_{x}x\right) \right],          \label{A:3c}
\\
\psi _{p_{x},-1,-1/2} &=&\frac{1}{2\sqrt{\varepsilon L}}
\begin{pmatrix}
\sqrt{\varepsilon -p_{x}} & -\sqrt{\varepsilon -p_{x}} \\
\sqrt{\varepsilon +p_{x}} & -\sqrt{\varepsilon +p_{x}}
\end{pmatrix}
\exp \left[ i\left( \varepsilon t+p_{x}x\right) \right],           \label{A:3d}
\end{eqnarray}
\end{subequations}
\end{widetext}
where $L$ is the normalized length.
In Eqs.~(\ref{A:3a})--(\ref{A:3d}),  the  first  (second) index of the matrices
corresponds to the spinor (isospinor) structure of the fermionic wave function.
The  wave  functions  $\psi_{p_{x}, \epsilon, I_{1}}$  satisfy  the  normalized
relations:
\begin{subequations}                                                \label{A:4}
\begin{eqnarray}
\bar{\psi}_{p_{x},\epsilon ,I_{1}}\gamma ^{\mu }\psi _{p_{x},\epsilon
,I_{1}} &=&\left( 1,\epsilon p_{x}\varepsilon ^{-1}\right) =
(1,\epsilon v_{x}),                                                \label{A:4a}
\\
\bar{\psi }_{\epsilon p_{x},\epsilon ,I_{1}}\gamma ^{\mu }\psi
_{\epsilon p_{x},\epsilon ,I_{1}} &=&\left( 1,p_{x}\varepsilon ^{-1}\right)
=(1,v_{x}),                                                        \label{A:4b}
\\
\bar{\psi }_{p_{x},\epsilon ,I_{1}}\psi _{p_{x},\epsilon ,I_{1}}
&=&\left( -1\right) ^{1/2-I_{1}}\epsilon M \varepsilon^{-1},       \label{A:4c}
\end{eqnarray}
\end{subequations}
where the normalized length $L$ is taken to be equal to unity.

Let us define the  spinor-isospinor  amplitude  $u_{p_{x}, \epsilon, I_{1}}$ of
the fermionic wave function $\psi_{p_{x},\epsilon,I_{1}}$ by the relation $\psi_{
p_{x},\epsilon,I_{1}}\equiv\left(2\varepsilon L\right)^{-1/2}u_{p_{x},\epsilon,
I_{1}} \exp\left[ - i \left( \epsilon \varepsilon t - p_{x} x \right) \right]$.
Then, the amplitudes $u_{p_{x},\epsilon,I_{1}}$ satisfy  the  orthogonality and
completeness relations:
\begin{subequations}                                                \label{A:5}
\begin{eqnarray}
u_{p_{x},\epsilon ,I_{1}}^{\dag }u_{p_{x},\epsilon ^{\prime },I_{1}^{\prime
}} &=&2\varepsilon \delta _{\epsilon ,\epsilon^{\prime }}\delta
_{I_{1},I_{1}^{\prime }},                                          \label{A:5a}
\\
\sum\limits_{\epsilon ,\,I_{1}}u_{p_{x},\epsilon ,I_{1}\left[ i,a\right]
}u_{p_{x},\epsilon ,I_{1}\left[j, b \right]}^{\dag} & = & 2 \varepsilon
\delta_{i,j}\delta_{a, b}                                          \label{A:5b}
\end{eqnarray}
\end{subequations}
for the Hermitian conjugate case, and
\begin{subequations}                                                \label{A:6}
\begin{flalign}
& \bar{u}_{\epsilon p_{x},\epsilon ,I_{1}}u_{\epsilon^{\prime
}p_{x},\epsilon ^{\prime},I_{1}^{\prime}} = 2 M\epsilon \left(-1\right)
^{1/2-I_{1}}\delta_{\epsilon ,\epsilon^{\prime }}\delta
_{I_{1},I_{1}^{\prime }},                                          \label{A:6a}
\\
& \sum\limits_{I_{1}}u_{\epsilon p_{x},\epsilon ,I_{1}\left[ i,a\right] }
\bar{u}_{\epsilon p_{x},\epsilon ,I_{1}\left[ j,b\right] } = \left\{
\gamma ^{\mu }p_{\mu }+\epsilon M\right\}_{i,j}\delta _{a,b},      \label{A:6b}
\\
& \sum\limits_{\epsilon ,\,I_{1}}\epsilon u_{\epsilon p_{x},\epsilon ,I_{1}
\left[ i,a\right] }\bar{u}_{\epsilon p_{x},\epsilon ,I_{1}\left[ j,b
\right] }  = 2 M \mathbb{\delta}_{i,j}\delta_{a,b}                 \label{A:6c}
\end{flalign}
\end{subequations}
for the Dirac conjugate case.
In Eqs.~(\ref{A:5b}), (\ref{A:6b}), and  (\ref{A:6c}), the first (second) index
in square brackets is the spinor (isospinor) one.

As for the Dirac equation (\ref{II:9}), the  free Dirac equation (\ref{A:1}) is
invariant      under      the     $C$,     $P$,     and   $T$   transformations
(\ref{II:15a})--(\ref{II:15c}).
However,   it  is  also  invariant  under  the  additional  variants  of  these
transformations:
\begin{subequations}                                                \label{A:7}
\begin{eqnarray}
\psi^{C^{\prime}} \left(t, x\right)  &=&\eta _{C^{\prime }}\gamma _{5}\otimes
\mathbb{I}\psi ^{\ast }\left( t, x\right),                         \label{A:7a}
\\
\psi^{C^{\prime \prime}} \left( t, x\right)  &=&\eta _{C^{\prime \prime }}
\mathbb{I}\otimes \tau _{2}\psi ^{\ast }\left( t,x\right),         \label{A:7b}
\\
\psi^{C^{\prime \prime \prime}} \left( t, x\right)  &=&
\eta _{C^{\prime \prime \prime }}
\mathbb{I}\otimes \tau _{3}\psi ^{\ast }\left( t,x\right),        \label{A:7bb}
\\
\psi^{P^{\prime}} \left( t,x\right)  &=&\eta _{P^{\prime }}\gamma ^{0}\otimes
\mathbb{I}\psi \left( t,-x\right),                                 \label{A:7c}
\\
\psi^{T^{\prime}} \left(t, x\right)  &=&\eta _{T^{\prime }}\gamma^{0}\otimes
\mathbb{I}\psi^{\ast }\left( -t, x\right).                         \label{A:7d}
\end{eqnarray}
\end{subequations}
In particular, it follows from Eqs.~(\ref{A:7b}) and (\ref{A:7bb}) that
\begin{subequations}                                                \label{A:8}
\begin{eqnarray}
\psi _{p_{x},\epsilon ,I_{1}} & \overset{C^{\prime \prime }}{\longrightarrow}
&\left(-1\right)^{\left(I_{1} + 1\right) /2}\psi_{-p_{x},-\epsilon ,-I_{1}},
                                                                   \label{A:8a}
\\
\psi_{p_{x},\epsilon ,I_{1}} &\overset{C^{\prime\prime\prime}}{\longrightarrow}
&\psi_{-p_{x},-\epsilon,-I_{1}},
                                                                   \label{A:8b}
\end{eqnarray}
\end{subequations}
where the phase  factors $\eta_{C^{\prime \prime}}$ and $\eta_{C^{\prime \prime
\prime}}$ are  taken equal to $-i$ and $1$, respectively.
It follows that in the  second  quantization  formalism, the negative frequency
fermionic wave function $\psi_{-p_{x}, -1, -I_{1}}$ can be used to describe the
antifermion having  the  momentum $p_{x}$  and  isospin $x$-projection $I_{1}$.

The free Hamiltonian (\ref{A:2}) commutes with the operator $T_{3} = \gamma_{5}
\!\otimes\!\tau_{3}$, which  determines  the type (diagonal or antidiagonal) of
the state.
At the same time,  the  operator $T_{3}$  does  not  commute  with  the isospin
operator $I_{1} = 2^{-1}\mathbb{I}\!\otimes\!\tau_{1}$.
It follows that the eigenstates of the operator $T_{3}$ are linear combinations
of the eigenstates of the isospin operator $I_{1}$:
\begin{subequations}                                                \label{A:9}
\begin{eqnarray}
\psi _{p_{x},1,\text{d}} &=&2^{-1/2}\left(\psi _{p_{x},1,1/2}-\psi
_{p_{x},1,-1/2}\right),                                            \label{A:9a}
  \\
\psi _{p_{x},1,\text{a}} &=&2^{-1/2}\left(\psi _{p_{x},1,1/2}+\psi
_{p_{x},1,-1/2}\right),                                            \label{A:9b}
  \\
\psi _{p_{x},-1,\text{d}} &=&2^{-1/2}\left(\psi _{p_{x},-1,1/2}-\psi
_{p_{x},-1,-1/2}\right),                                           \label{A:9c}
  \\
\psi _{p_{x},-1,\text{a}} &=&2^{-1/2}\left(\psi _{p_{x},-1,1/2}+\psi
_{p_{x},-1,-1/2}\right).                                           \label{A:9d}
\end{eqnarray}
\end{subequations}

\section{Energy of a bound fermionic (antifermionic) state in the vicinity
of the transition to the continuum}

In this appendix, we explain the behavior of the curves $\left\vert\varepsilon_{
n}\right\vert/M$ expressed by Eq.~(\ref{IV:5}).
Since  for  the  diagonal  and   antidiagonal   cases  the  curves  $\left\vert
\varepsilon_{n}\left( M\right)\right\vert$ are the same, we shall consider only
the diagonal case.
As $M$ tends to $M_{n}$ from  the  right, the curve $\left\vert \varepsilon_{n}
\left(M\right) \right\vert/M$ tends to unity.
It follows that in this case, the parameter $\kappa=\left(M^{2}-\varepsilon^{2}
\right)^{1/2}$ tends to zero.

Let  us  replace  the   independent  variable  $x$ in the differential equation
(\ref{III:8}) by $\xi = \kappa x$.
After this replacement, the differential equation takes the form
\begin{eqnarray}
&&\psi _{11}^{\prime \prime }(\xi )+2im\kappa^{-1}\text{sech}\left(m\kappa
^{-1}\xi \right) \psi_{11}^{\prime}(\xi)                          \nonumber
\\
&&-\left( 1-2\varepsilon m\kappa ^{-2}\text{sech}\left( m\kappa ^{-1}\xi
\right) \right) \psi _{11}(\xi ) = 0.                               \label{B:1}
\end{eqnarray}
When the dimensionless variable  $\left\vert \xi \right\vert\gg \kappa m^{-1}$,
the function $\text{sech}\left(m\kappa ^{-1}\xi \right)$ exponentially tends to
zero.
In this case, we can neglect the  corresponding  terms  in  Eq.~(\ref{B:1}) and
find that
\begin{equation}
\psi_{11}(\xi)\propto \exp\left(-\left\vert \xi \right\vert \right) \label{B:2}
\end{equation}
when $\left\vert \xi \right\vert\gg \kappa m^{-1}$.
Next we turn to the region  of  $\left\vert \xi \right\vert \ll \kappa m^{-1}$,
where the function $\text{sech}\left(m \kappa^{-1}\xi\right)$ can  be set equal
to unity. 
Using this fact, we find  an  approximate  solution  to  Eq.~(\ref{B:1}) in the
region of $\left\vert \xi \right\vert \ll \kappa m^{-1}$:
\begin{equation}
\psi_{11}(\xi ) \propto \exp \left(-i\kappa ^{-1}\tau \xi \right),  \label{B:3}
\end{equation}
where the parameter
\begin{equation}
\tau=m+\epsilon\left(m^{2}+2 m\varepsilon - \kappa^{2}\right)^{1/2} \label{B:4}
\end{equation}
and $\epsilon = \pm 1$.

Now we need to match solutions (\ref{B:2}) and (\ref{B:3}) at $\xi = \kappa m^{
-1}$.
Since,  for  $\left\vert  \xi \right\vert \ll \kappa  m^{-1}$,  Eq.~(\ref{B:1})
contains the imaginary coefficient  $2i m \kappa^{-1}$, solution (\ref{B:3}) is
essentially complex.
The matching condition must be  satisfied for both the real and imaginary parts
of solution (\ref{B:3}).
Let us consider the real part of Eq.~(\ref{B:3}):
\begin{equation}
\text{Re}\left[\psi_{11}(\xi)\right]\propto\cos\left(\kappa^{-1}\tau\xi\right).
                                                                    \label{B:5}
\end{equation}
By equating the logarithmic derivatives  of Eqs.~(\ref{B:2}) and (\ref{B:5}) at
$\xi = \kappa m^{-1}$, we arrive at the transcendental equation
\begin{equation}
\tau \tan \left( m^{-1}\tau \right) = \kappa.                       \label{B:6}
\end{equation}
From Eq.~(\ref{B:6})  it  follows  that  $\tan \left( m^{-1}\tau \right)$  must
tend to zero together with $\kappa$.
This, in turn, implies  that  the  combination  $m^{-1}\tau$  can be written as
\begin{equation}
m^{-1}\tau = \pi n + \Delta,                                        \label{B:7}
\end{equation}
where $n$  is  a  nonnegative  integer  and  $\Delta \rightarrow 0$  as $\kappa
\rightarrow 0$.
Combining Eqs.~(\ref{B:4}) and (\ref{B:7}), we find that
\begin{equation}
\left\vert \varepsilon _{n}\right\vert \underset{\kappa \rightarrow 0}
{\longrightarrow }M_{n}=m\pi n\left( n\pi /2-1\right),              \label{B:8}
\end{equation}
where $\varepsilon_{n}$  is  the energy of the $n$-th fermionic (antifermionic)
bound state.
We see that for large $n$, the value  of  the  fermion mass at which the $n$-th
fermionic  (antifermionic)  bound  state  arises  from  the  continuum  becomes
approximately proportional to $n^{2}$.
Note that this fact is in  accordance  with Eqs.~(\ref{IV:2}) and (\ref{IV:3}).

By  substituting   Eq.~(\ref{B:7})   into   Eq.~(\ref{B:6}),   we   obtain  the
transcendental equation  in  terms  of  the  parameters  $\Delta$ and $\kappa$:
\begin{equation}
m\left(\pi n + \Delta \right) \tan \left(\Delta \right) = \kappa.   \label{B:9}
\end{equation}
We  now  study  the  behavior  of   the   curve  $\left\vert \varepsilon_{n}(M)
\right\vert$ in the neighborhood  of  the  fermion  mass  $M_{n}$,  where  both
$\Delta$ and $\kappa$ tend to zero.
Expanding the right-hand side  of  Eq. (\ref{B:9})  in  terms  of  $\Delta$ and
keeping the first expansion terms, we obtain the equation
\begin{equation}
m n\pi \Delta \left(1 - \delta_{n0}\right) + m \Delta^{2}\delta_{n0} = \kappa.
                                                                   \label{B:10}
\end{equation}
In Eq.~(\ref{B:10}),  we  can  express  both  $\Delta$ and $\kappa$ in terms of
$\varepsilon$ and $M$, using  Eqs.~(\ref{B:4}), (\ref{B:7}), and the definition
$\kappa = \left( M^{2}-\varepsilon^{2}\right)^{1/2}$.
In the neighborhood  of  $M_{n}$ (defined  in  Eq.~(\ref{B:8})),  the variables
$\varepsilon$ and $M$ can be written as
\begin{subequations}                                               \label{B:11}
\begin{eqnarray}
\varepsilon  & = &M_{n}+\Delta \varepsilon,                       \label{B:11a}
\\
M & = &M_{n}+\Delta M.                                            \label{B:11b}
\end{eqnarray}
\end{subequations}
As a result, we obtain  a  cumbersome  expression  implicitly  defining $\Delta
\varepsilon$ as a function of $\Delta M$ in the neighborhood of $M_{n}$:
\begin{equation}
F\left( m,n,\Delta M,\Delta \varepsilon \right) = 0.               \label{B:12}
\end{equation}
The value of $\Delta \varepsilon$ should  vanish along with that of $\Delta M$,
and  therefore  Eq.~(\ref{B:12})  must  be  satisfied  identically when $\Delta
\varepsilon$ and $\Delta M$ vanish.
This can be used to determine  the  sign  factor $\epsilon$ in Eq.~(\ref{B:4}):
\begin{equation}
\epsilon = 1 - 2 \delta_{n0},                                      \label{B:13}
\end{equation}
where it is understood that  in  Eq.~(\ref{B:4}),  the  principal  value of the
square root is used.

\begin{figure}[tbp]
\includegraphics[width=0.5\textwidth]{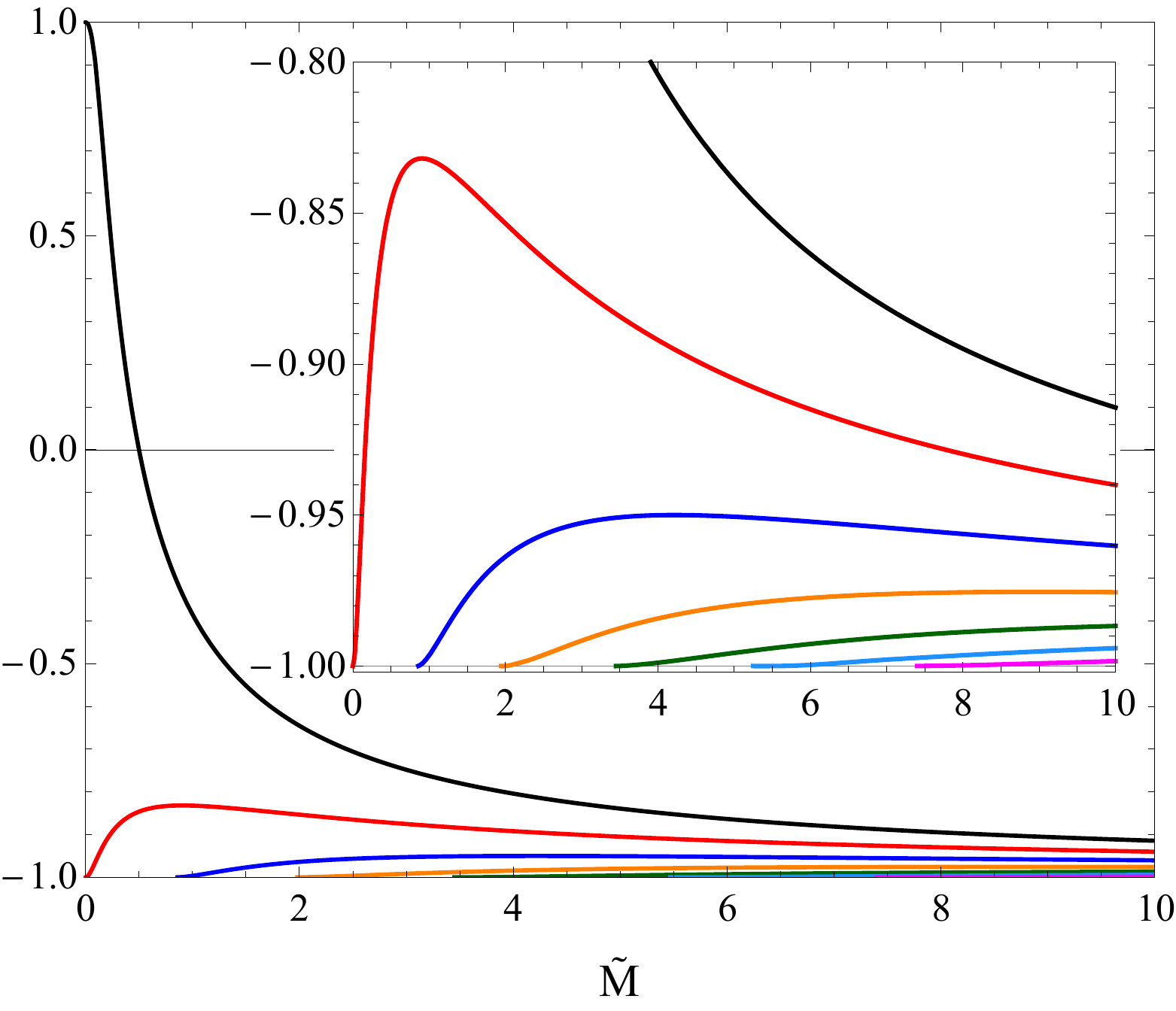}
\caption{\label{fig7}   Dependence of the ratio $\tilde{\varepsilon}/\tilde{M}$
for the antidiagonal bound states  of the Dirac Hamiltonian on the fermion mass
$\tilde{M}$.}
\end{figure}

Using expression (\ref{B:12}), treating $\Delta \varepsilon$  as  a function of
$\Delta M$, and applying the rules for differentiation of an implicit function,
we obtain sequentially:
\begin{subequations}                                               \label{B:14}
\begin{eqnarray}
\Delta \varepsilon \left( 0\right)  &=&0,                         \label{B:14a}
 \\
\Delta \varepsilon ^{\prime }\left( 0\right)  &=&1,               \label{B:14b}
 \\
\Delta \varepsilon ^{\prime \prime }\left( 0\right)  &=&-\frac{2\pi n}{
m\left( \pi n-2\right) \left( \pi n-1\right) ^{2}},               \label{B:14c}
 \\
\Delta \varepsilon ^{\prime \prime \prime }\left( 0\right)  &=&\frac{6}{m^{2}
}\frac{-2+4\pi n-\pi ^{2}n^{2}+\pi ^{2}n^{3}}{\left( \pi n-2\right)
^{2}\left( \pi n-1\right)^{3}}.                                   \label{B:14d}
\end{eqnarray}
\end{subequations}
for $n = 1, 2, 3, \ldots ,$ and
\begin{subequations}                                               \label{B:15}
\begin{eqnarray}
\Delta \varepsilon \left( 0\right)  &=&0,                         \label{B:15a}
 \\
\Delta \varepsilon ^{\prime }\left( 0\right)  &=&\pm 1,           \label{B:15b}
 \\
\Delta \varepsilon ^{\prime \prime }\left( 0\right)  &=&0,        \label{B:15c}
 \\
\Delta \varepsilon^{\prime \prime \prime}\left(0\right) &=&\mp 3m^{-2}
                                                                  \label{B:15d}
\end{eqnarray}
\end{subequations}
for $n = 0$.
The two signs in  Eqs.~(\ref{B:15b})  and  (\ref{B:15d})  correspond to the two
curves starting at $\tilde{M} = 0$ in Fig.~\ref{fig7}.

It follows  from  Eqs.~(\ref{B:14})  and  (\ref{B:15}) that $\left \vert \Delta
\varepsilon^{\prime}\left( 0 \right) \right \vert$  is exactly equal to one and
that $\Delta \varepsilon^{\prime \prime}\left(0\right)$  vanishes when $n = 0$.
We can show that this behavior of the function $\Delta \varepsilon \left(\Delta
M \right)$ corresponds to Eq.~(\ref{IV:5}).
To do this, we suppose that in the neighborhood of $M_{n}$, the function $\left
\vert \varepsilon_{n}\right\vert/M$ has the form
\begin{eqnarray}
\frac{\left\vert \varepsilon _{n}\right\vert }{M} &=&1-\alpha _{n}\left(
M - M_{n}\right) - \beta_{n}\left(M - M_{n}\right)^{2}       \nonumber
 \\
&&-\gamma_{n}\left(M - M_{n}\right)^{3} + O\left[\left(M - M_{n}\right)^{4}
\right],                                                           \label{B:16}
\end{eqnarray}
where $\alpha_{n}$,  $\beta_{n}$,  and $\gamma_{n}$  are constant coefficients.
Rewriting  Eq.~(\ref{B:16})  in   terms   of  $\Delta \varepsilon = \left \vert
\varepsilon_{n}\right\vert - M_{n}$  and  $\Delta M = M - M_{n}$, we obtain the
expression
\begin{eqnarray}
\Delta \varepsilon  &=&\left( 1-M_{n}\alpha _{n}\right) \Delta M-\left(
\alpha _{n}+M_{n}\beta _{n}\right) \Delta M^{2}       \nonumber
 \\
&&-\left(\beta_{n} + M_{n}\gamma_{n}\right)\Delta M^{3} +
O\left(\Delta M^{4}\right).                                        \label{B:17}
\end{eqnarray}
From Eqs.~(\ref{B:14}) and (\ref{B:15}) it  follows  that  in Eq.~(\ref{B:17}),
the coefficient $\alpha_{n}$ vanishes,  whereas  the coefficient $\beta_{n} \ne
0$ and is positive for all $n$.
Under these conditions,  Eqs.~(\ref{IV:5})  and (\ref{B:16}) become equivalent.
Note also that in  Eq.~(\ref{B:17}), the coefficient at $\Delta M^{2}$ vanishes
when $n = 0$, in accordance with Eq.~(\ref{B:15c}).

We conclude that the use of the rather rough approximation allows us to explain
the behavior of the curves  in  Fig.~\ref{fig7} at a qualitative level.

\bibliography{article}
\clearpage

\clearpage

\clearpage

\end{document}